\documentclass[twocolumn,twocolappendix]{aastex7}

\makeatletter
\renewcommand\paragraph{\@startsection{paragraph}{4}{\z@}%
  {1.5ex \@plus 0.5ex \@minus 0.2ex}%
  {-1em}%
  {\normalfont\normalsize\itshape}}
\makeatother

\usepackage{amsmath}

\newcommand\gal{\texttt{galacticus}}
\newcommand\kms{km$\,$s$^{-1}$}
\newcommand\msun{M_\mathrm{\odot}}
\newcommand\msub{M_\mathrm{sub}}

\newcommand\mthresh{5 \times 10^5}
\newcommand\vkick{v_\mathrm{kick}}
\newcommand\nstream{24}

\begin{document}

\title{Which Milky Way Streams Make the Best Dark Matter Detectors?}
\shorttitle{Which Streams Make the Best DM Detectors?}
\shortauthors{Chandra et al.}

\author[0000-0002-0572-8012]{Vedant Chandra}
\altaffiliation{NASA Hubble Fellow}
\affiliation{Center for Astrophysics $\mid$ Harvard \& Smithsonian, 60 Garden St, Cambridge, MA 02138, USA}
\affiliation{MIT Kavli Institute for Astrophysics and Space Research, 70 Vassar St, Cambridge, MA 02139, USA}
\email{chandra@mit.edu}

\author[0000-0002-1590-8551]{Charlie~Conroy}
\affiliation{Center for Astrophysics $\mid$ Harvard \& Smithsonian, 60 Garden St, Cambridge, MA 02138, USA}
\email{cconroy@cfa.harvard.edu}

\author[0000-0003-2229-011X]{Risa~H.~Wechsler}
\affiliation{Kavli Institute for Particle Astrophysics and Cosmology, Stanford University, Stanford, CA 94305, USA}
\affiliation{Department of Physics, Stanford University, Stanford, CA 94305, USA}
\affiliation{SLAC National Accelerator Laboratory, Menlo Park, CA 94025, USA}
\email{rwechsler@stanford.edu}

\author[0000-0002-7846-9787]{Ana~Bonaca}
\affiliation{The Observatories of the Carnegie Institution for Science, 813 Santa Barbara Street, Pasadena, CA 91101, USA}
\email{abonaca@carnegiescience.edu}

\author[0000-0001-5501-6008]{Andrew~Benson}
\affiliation{The Observatories of the Carnegie Institution for Science, 813 Santa Barbara Street, Pasadena, CA 91101, USA}
\email{abenson@carnegiescience.edu}

\author[0000-0001-8042-5794]{Jacob~Nibauer}
\affiliation{Department of Astrophysical Sciences, Princeton University, 4 Ivy Lane, Princeton, NJ 08544, USA}
\affiliation{Center for Astrophysics $\mid$ Harvard \& Smithsonian, 60 Garden St, Cambridge, MA 02138, USA}
\email{jnibauer@cfa.harvard.edu}

\author[0000-0002-4746-2128]{Tara~Dacunha}
\affiliation{Kavli Institute for Particle Astrophysics and Cosmology, Stanford University, Stanford, CA 94305, USA}
\affiliation{Department of Physics, Stanford University, Stanford, CA 94305, USA}
\affiliation{SLAC National Accelerator Laboratory, Menlo Park, CA 94025, USA}
\email{tdacunha@stanford.edu}

\author[0000-0001-9863-5394]{Philip~Mansfield}
\affiliation{Kavli Institute for Particle Astrophysics and Cosmology, Stanford University, Stanford, CA 94305, USA}
\email{mansfield.astro@gmail.com}

\author{Paul~Menker}
\affiliation{Department of Physics and Astronomy, University of Southern California, 3551 Trousdale Pkwy, Los Angeles, CA 90089, USA}
\affiliation{The Observatories of the Carnegie Institution for Science, 813 Santa Barbara Street, Pasadena, CA 91101, USA}
\email{pmenker@usc.edu}

\author[0009-0005-1914-974X]{Anya~Phillips}
\affiliation{Center for Astrophysics $\mid$ Harvard \& Smithsonian, 60 Garden St, Cambridge, MA 02138, USA}
\email{anya.phillips@cfa.harvard.edu}

\correspondingauthor{Vedant Chandra}
\email{chandra@mit.edu}

\begin{abstract}

\noindent
The abundance and properties of low-mass dark matter subhalos can be inferred by measuring the density and velocity structure of stellar streams, testing physical models of dark matter. 
Which streams make the most promising dark matter detectors, and which subhalos would they encounter?
We use the semi-analytic model \texttt{galacticus} to simulate the Milky Way's dark subhalos, including tidal stripping under the influence of the galactic disk.
Deriving stream initial conditions that best match present-day observations, we forecast subhalo encounters for 24 globular cluster streams around the Milky Way.
Over their lifetimes, the 10 most perturbed streams are predicted to each experience $\approx 2$--$12$ significant subhalo encounters with velocity kicks $v_\mathrm{kick} > 0.1$\,km\,s$^{-1}$.
Jet, AAU, GD-1, and Palomar~5 lead the list, with Jet encountering twice the subhalos of any other stream. 
The subhalos that encounter streams have been tidally stripped to $1$--$20\%$ of their peak mass and are twice as compact as previous works have assumed. 
A spectroscopic survey of streams with sub-km\,s$^{-1}$ precision could reveal the influence of dark subhalos below the threshold of galaxy formation. 
Predicting the abundance and scatter of low-mass subhalos that reach the inner galaxy remains a dominant uncertainty in connecting stellar streams to models of dark matter. 

\end{abstract}

\section{Introduction} \label{sec:intro}

Although dark matter (DM) plays a defining role in our cosmological paradigm---outweighing baryonic matter by almost a factor of five---the true particle nature of DM remains mysterious. 
Theoretical models of particle DM span many orders of magnitude in mass, and the nature of DM's interactions with itself and with the Standard Model remains unknown \citep[e.g.,][]{Bertone2005, Bertone2018}. 
Numerous experiments are underway to directly detect DM particles in laboratories, and current and upcoming facilities are searching for gamma-ray signatures of DM self-annihilation \citep{Schumann2019}. 

In the standard paradigm of ``cold'' and collisionless DM (CDM), gravitational collapse is expected to create small-scale substructure in the form of subhalos down to very small mass scales \citep[e.g.,][]{Klypin1999, Moore1999}. 
Departures from the CDM model come in several flavors. 
Warm dark matter (WDM) models predict that DM particles retain significant thermal motion when they freeze out in the early Universe \citep{Colombi1996, Bode2001, Schutz2026}. 
This free-streaming motion inhibits the growth of structure below a characteristic mass scale that depends on the mass of the WDM particle candidate, resulting in a low-mass cutoff in the subhalo mass function (SHMF; e.g., \citealt{Lovell2020, Nadler2025a}). 
Other models like fuzzy dark matter (FDM; \citealt{Hu2000}) and self-interacting dark matter (SIDM; \citealt{Spergel2000}) can also alter the SHMF, and additionally modify the internal density structure of subhalos. 
SIDM in particular predicts a population of both cored and core-collapsed subhalos, increasing the diversity of subhalo densities compared to CDM \citep{Spergel2000, Bechtol2022}. 

Measuring the abundance and characteristics of low-mass DM subhalos remains the most promising astrophysical method to understand the true particle nature of DM \citep[e.g.,][]{Bertone2018}. 
In external galaxies, subtle perturbations to the images of strong gravitational lenses enable the inference of the underlying population of DM subhalos around a lens \citep[see][for a review]{Vegetti2024}. 
These methods have been used to place lower limits on the amount of DM contained in subhalos, suggesting that subhalos persist down to peak masses (prior to tidal stripping) $M_\mathrm{peak} \gtrsim 10^8\,M_\odot$ \citep{Mao1998, Gannon2025, Gilman2025, Gilman2026, Nierenberg2026}. 
Around the Milky Way (MW), over 60 DM subhalos have revealed their presence by hosting detectable dwarf galaxies \citep[e.g.,][]{Simon2019}. 
Matching the population of dwarf galaxies to the halos in which they likely reside already places stringent limits on alternative models of dark matter, implying that the faintest known dwarfs occupy halos with peak masses $M_\mathrm{peak} \gtrsim 10^8\,M_\odot$ \citep[e.g.,][]{Nadler2020, Nadler2021, Nadler2021unified}. 
Differentiating DM models whose predictions diverge below $M_\mathrm{peak} \lesssim 10^7$--$10^8\,M_\odot$ would require the detection and characterization of truly dark, starless subhalos around the MW \citep[e.g.,][]{Nadler2025, Nadler2026review}. 

The advent of wide-field photometric and astrometric surveys \citep[e.g.,][]{York2000, GaiaCollaboration2016, DarkEnergySurveyCollaboration2016, Dey2019} has led to the discovery of over a hundred stellar streams that encircle the MW \citep[e.g.,][]{Majewski2003, Belokurov2006b, Grillmair2006, Grillmair2009, Newberg2009, Shipp2018, Ibata2019, Helmi2020, Malhan2022a, Mateu2022, Bonaca2025}. 
These are dwarf galaxies or globular clusters that have dissolved (or are in the process of doing so) due to the tidal influence of the MW. 
In the absence of external perturbations, stream stars closely trace the orbits of their progenitors. 
It has been recognized that this property of streams---particularly cold, thin streams originating from globular clusters---makes them promising detectors of gravitational perturbers like DM subhalos \citep[e.g.,][]{Johnston2002, Ibata2002, Yoon2011, Carlberg2012, Erkal2015a, Erkal2015, Erkal2016a, Bovy2017, Bonaca2018, Bonaca2019, Bonaca2025, Pelissier2026}. 
Since streams accumulate the entire history of their DM perturbations---including those from starless subhalos---they may hold the key to advancing our understanding of particle DM \citep[e.g.,][]{DrlicaWagner2019, Banik2019, Banik2021streams, AnauMontel2026, Nibauer2026, Nadler2026review, ViaCollaboration2026}. 

Several recent works have estimated the influence of DM subhalos on the densities and kinematics of stellar streams. 
\cite{Erkal2016a} developed a simplified analytic formulation for the flux of subhalos encountered by a given stellar stream; focusing on strong gap-forming encounters, they predict $\mathcal{O}(1)$ significant gaps in a stream like Palomar~5.  
\cite{Barry2023} and \cite{Arora2024} predict stream--subhalo encounters in the FIRE hydrodynamical simulations, finding that the GD-1 stream undergoes $\approx 5$ close subhalo encounters per Gyr, and that the influence of a satellite like the Large Magellanic Cloud (LMC) imprints spatial asymmetries in the subhalo population. 
\cite{Adams2025} and \cite{Menker2026} predict gap-forming encounters for GD-1 and Palomar 5, respectively, using semi-analytic models for the subhalo distribution \citep{Benson2012, Jiang2021}, again finding that these streams should have $\mathcal{O}(1)$ substantial gaps, with numerous additional weak perturbations. 
\cite{Lu2026} and \cite{Lu2026b} couple the \cite{Erkal2016a} subhalo model to an observational model for real MW streams, quantifying the ability to detect individual subhalo encounters with photometric and radial velocity (RV) information. 
\cite{Pelissier2026} predict the ability of density information from the Vera Rubin Observatory's Legacy Survey of Space and Time \citep[LSST;][]{Ivezic2019} to detect stream gaps induced by subhalos, highlighting the influence of observational systematics, and motivating additional information like proper motions and RVs to characterize low-mass subhalos.

We are entering a new era of wide-field surveys targeting MW streams with an eye toward measuring the DM substructure in the MW. 
The Rubin Observatory's LSST, in concert with upcoming data releases from the \textit{Gaia} space observatory \citep{GaiaCollaboration2016, GaiaCollaboration2022}, will greatly enhance the detection and morphological analysis of stellar streams \citep[e.g.,][]{DrlicaWagner2019, Bonaca2025}. 
Furthermore, ongoing and upcoming spectroscopic surveys like DESI \citep{DESICollaboration2016}, SDSS-V \citep{Kollmeier2025}, 4MOST \citep{deJong2019}, S$^5$ \citep{Li2019}, Via \citep{ViaCollaboration2026}, and Spec-S5 \citep{Besuner2025} will map streams across the sky, confirming the membership of stream stars, and measuring the kinematic structure of streams \citep[e.g.,][]{Carlberg2025, Jarvis2026, Nibauer2026}.

Which of the MW's streams make the best DM detectors, and what could a spectroscopic survey of these streams teach us about DM? 
Answering this question requires a robust model for the evolution of stellar streams over their lifetimes, as well as a cosmologically realistic population of MW subhalos. 
In this work, we consider the entire ensemble of promising globular cluster streams known with \textit{Gaia} data \citep{Bonaca2025}, and derive stream simulations that model the evolution of each stream over its lifetime. 
We couple these stream simulations to a realistic simulation of MW subhalos predicted by the semi-analytic model \gal{} \citep{Benson2012}, modeling multiple realizations of the subhalo evolution including the gravitational influence of the MW's disk. 
We use this simulation to evaluate the characteristics of subhalo encounters experienced by each stream, identify the most promising DM detectors, and characterize the DM subhalo population that they are sensitive to. 

\begin{figure}
    \centering
    \includegraphics[width=\linewidth]{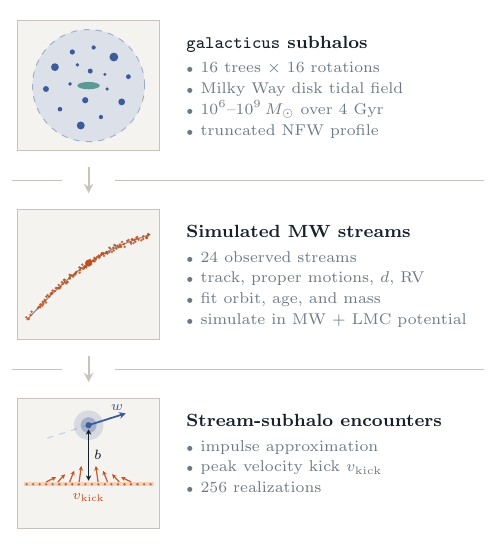}
    \caption{\textbf{Schematic of the three simulation stages presented in this work.} Cosmologically realistic subhalo populations are generated with the semi-analytic model \gal{}, and stellar stream models are fit to match the  evolution of \nstream{} known MW streams. The resulting stream--subhalo encounters are measured and characterized.}
    \label{fig:schematic}
\end{figure}

A schematic summary of our methodology is presented in Figure~\ref{fig:schematic}.
$\S$\ref{sec:dm_model} describes our subhalo simulations with \gal{}, including an evaluation of the underlying subhalo distribution, and modeling the individual density profiles of subhalos. 
We present our stellar stream models in $\S$\ref{sec:streams}, selecting candidate streams with \textit{Gaia} data and deriving initial conditions for their progenitors that best match their present-day structure and kinematics. 
Our results are presented in $\S$\ref{sec:results}: quantitative forecasts for the subhalo encounters expected for each stream, the properties of the stream-detectable population of subhalos, and simple estimates for how these predictions would differ in alternative models of DM. 
We contextualize and discuss our results in $\S$\ref{sec:discuss}, and list our conclusions in $\S$\ref{sec:conclusions}. 

\section{Modeling Dark Matter Subhalos}\label{sec:dm_model}

In this section, we describe our methodology to construct a statistically plausible population of MW subhalos, and to determine progenitor orbits for the known stellar streams around the MW. 
With these simulations in hand, we calculate the abundance and characteristics of stream--subhalo encounters. 

\subsection{Subhalo Distributions in \texttt{galacticus}}\label{sec:subdist}

The first ingredient in our simulation is the underlying population of MW subhalos. 
The highest-fidelity forecast would be a suite of $N$-body simulations with sufficient mass resolution to reliably trace subhalos at our mass scale of interest. 
However, the field has not yet produced full suites of simulations that have both $\mathcal{O}(10)$ statistical realizations of the MW and reliably resolved subhalos down to the $\sim 10^6\,M_\odot$ regime relevant for stream perturbations \citep[e.g.,][]{Nadler2023, Mansfield2024}. 
%However, even the most detailed MW zoom simulations cannot reliably resolve subhalos down to the $\sim 10^6\,M_\odot$ regime relevant for stream perturbations \citep[e.g.,][]{Nadler2023, Mansfield2024}. 
%Furthermore, due to computational expense, these simulations only exist for $\mathcal{O}(10)$ statistical realizations of the MW. 
Therefore, it is challenging to marginalize over cosmological variance---and understand the source and magnitude of the halo--halo scatter---using $N$-body simulations alone. 
 
In this work, we model the evolution of DM subhalos using the semi-analytical model \gal{}, including the tidal effects of a stellar disk on subhalo disruption \citep{Benson2012, Pullen2014, Yang2020, Benson2022, Menker2026}.
For a target halo mass at $z = 0$, \gal{} constructs a halo merger tree and simulates it forward in time, modeling the orbital evolution of subhalos including mergers, tidal stripping, and tidal heating \citep{Pullen2014, Jiang2015, Benson2020, Benson2022}. 
Halos are originally assumed to follow \citet[][NFW]{Navarro1997} profiles at infall, following the concentration model of \cite{Johnson2021concentrations}, and are subsequently evolved with a semi-analytic orbital and tidal model. 
The \gal{} prescriptions have been tuned to match mass function and tidal stripping predictions from $N$-body simulations at both cosmological and subhalo-level scales \citep{Yang2020, Nadler2023}. 
We use the public version of \gal{}, accessed via GitHub\footnote{\url{https://github.com/galacticusorg/galacticus}} on 2025-05-15.

\begin{figure}
    \centering
    \includegraphics[width=\columnwidth]{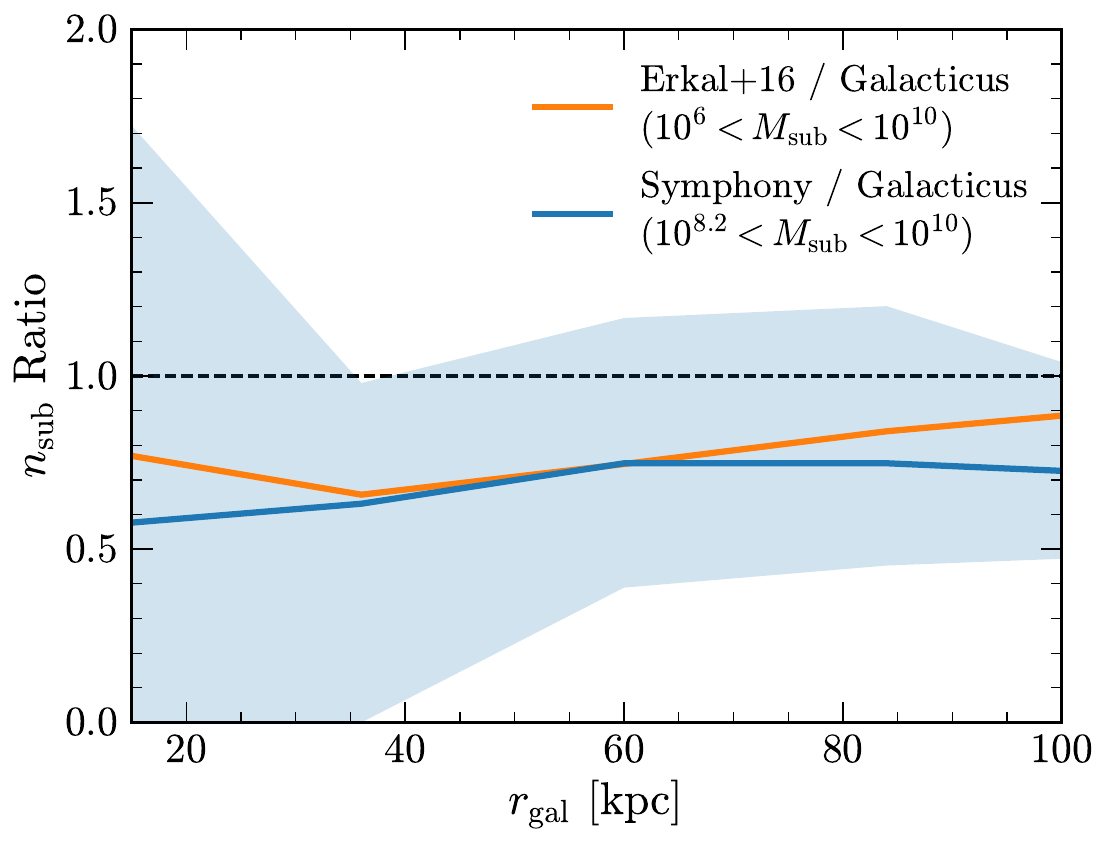}
    \includegraphics[width=\columnwidth]{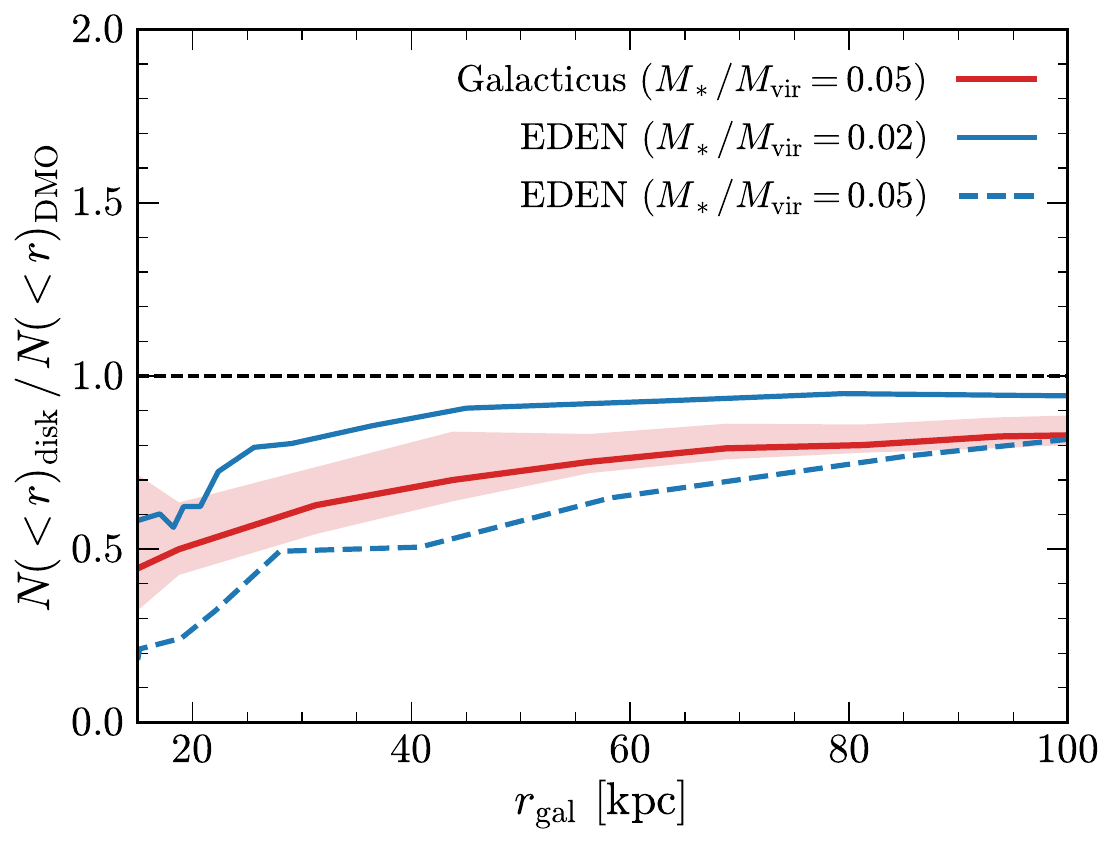}
    \caption{\textbf{Radial distribution of subhalos in our \gal{} simulation. Top:} The number density ratio of subhalos as a function of Galactocentric radius is shown for a DM-only model with no disk potential, compared to values from the \texttt{Symphony} $N$-body simulation \citep{Nadler2023} and the \cite{Erkal2016a} model commonly used to predict stream--subhalo interactions.
    \textbf{Bottom:} Cumulative subhalo abundance ratio between \gal{} simulations with and without a MW disk potential, highlighting the suppression of subhalos in the inner regions of the galaxy. Corresponding curves from the \texttt{Symphony}-EDEN $N$-body simulation are also shown for two disk masses \citep{Wang2025}. 
    }
    \label{fig:nsub_comparison}
    \label{fig:nsub_disk}
\end{figure}

To facilitate a comparison to previous works, we first conduct DM-only \gal{} runs on $64$ host halos and examine the resulting distributions of subhalos at $z = 0$. 
We adopt a host halo mass of $1.2 \times 10^{12}\,M_\odot$ for this test, matching the median MW-like host mass in the \texttt{Symphony} suite of $N$-body simulations \citep{Nadler2023}. 
Figure~\ref{fig:nsub_comparison} shows the mean number density of subhalos as a function of Galactocentric radius, as a ratio compared to \texttt{Symphony}, using subhalos detected by the \texttt{Symfind} tracking algorithm \citep{Mansfield2024}.
We also show the analytic relation used by the stream--subhalo forecasts of \cite{Erkal2016a} --- rescaled to the \texttt{Symphony} host mass --- which is tuned to the Aquarius suite of $N$-body simulations \citep{Springel2008}.
For the purposes of this comparison, we use a DM-only \gal{} run, and do not apply the disk-depletion factor from \cite{Erkal2016a}, which is itself based on \cite{DOnghia2010} (our actual simulation does use an analytic disk potential to self-consistently boost the tidal stripping of subhalos, as described later in this section). 
For the \texttt{Symphony} comparison, we restrict to subhalos with $10^{8.2} < \msub/M_\odot < 10^{10}$, the regime in which \texttt{Symphony} subhalos are well-resolved \citep{Nadler2023, Mansfield2024}. 
In the inner regions of the galaxy, \gal{} predicts up to $\approx 1.5\times$ more surviving subhalos than \cite{Erkal2016a} or the median \texttt{Symphony} halo, albeit within the $1\sigma$ envelope of the \texttt{Symphony} suite. 

The discrepancy between the inner galaxy abundance of subhalos predicted by \gal{} and \texttt{Symphony} was pointed out by \cite{Nadler2023}, and could be partially due to resolution effects in the \texttt{Symphony} suite causing an under-counting of surviving subhalos, particularly in the inner regions of the host halo.
This could also be true of the Aquarius suite, on which the model used by \cite{Erkal2016a} is tuned, particularly due to their older halo-finding algorithm potentially losing subhalos in the inner regions of the galaxy \citep{Mansfield2024}. 
However, our version of \gal{} is tuned to the Caterpillar suite of $N$-body simulations \citep{Griffen2016, Yang2020}, whose subhalo mass functions are known to differ from \texttt{Symphony} at the $\approx 20\%$ level without a known explanation \citep[see Appendix~E in][]{Nadler2023}.
Furthermore, the mass loss prescription in \gal{} was not tuned to match the radial number density profile for MW-mass hosts, and may require further adjustment.
We proceed with the \gal{} predictions for the rest of this work, noting that the absolute normalization of subhalos in the inner regions of MW-mass hosts warrants further study, potentially with higher-resolution $N$-body simulations capable of resolving $\sim 10^6$--$ 10^7\,M_\odot$ subhalos in the inner galaxy, over a number of realizations consistent with the properties of the Milky Way. Even a small number of such simulations would help calibrate semi-analytic models like \gal{} to produce statistical realizations of the low-mass subhalo population \citep[e.g.,][]{Dacunha2026}.

In the inner regions of the MW's DM halo, the stellar disk dominates the gravitational potential, leading to an enhancement in the tidal stripping and destruction of subhalos \citep[e.g.,][]{GarrisonKimmel2017, Green2022, Wang2025}. 
\cite{Wang2025} embedded disk potentials in \texttt{Symphony} simulations, finding that the abundance of subhalos within $r_\mathrm{gal} \lesssim 20$~kpc can be reduced to $\approx 50\%$ of the DM-only subhalo abundance. 
Motivated by this, we implement an exponential disk potential in \gal{} which self-consistently influences the tidal evolution of subhalos. 
The disk evolution is governed by the \texttt{UniverseMachine} \citep{Behroozi2019} stellar mass prescriptions built into \gal{}, with the $z=0$ disk mass constrained to reach $5 \times 10^{10}\,\msun$ to match MW observations \citep{Bland-Hawthorn2016}. 
Figure~\ref{fig:nsub_disk} illustrates the influence of this disk potential. 
The abundance of subhalos near the disk is reduced to $\approx 50\%$ of the DM-only subhalo abundance with a strong radial dependence.
This is broadly consistent with the results of \cite{Wang2025}, although at a fixed disk-to-host mass ratio the \cite{Wang2025} simulations predict stronger subhalo suppression than \gal{}. 
Figure~\ref{fig:nsub_disk} also suggests that the radius--independent factor-of-three disk-induced subhalo suppression assumed in the \cite{Erkal2016a} model (based on \citealt{DOnghia2010}) is too strong, with the \gal{} and \texttt{Symphony} suppression only reaching that value in the very inner $\lesssim 10$~kpc of the galaxy. 
Over the radial range of subhalos probed by MW streams (see Figure~\ref{fig:rgal_dist}), $55$--$85\%$ of DM-only subhalos survive the influence of the disk, with the surviving fraction rising as a function of Galactocentric radius. 
Our \gal{} model self-consistently models the disk's influence on each subhalo based on its specific orbit, for example heavily stripping subhalos on eccentric orbits with small pericenters. 

\subsection{Subhalo Density Profiles}

The gravitational interaction between DM subhalos and stellar streams is softened and altered by the extended mass distributions of the subhalos. 
It is therefore important to adopt realistic density distributions for the individual subhalos. 
The \gal{} prescription for tidal heating and tidal stripping produces numerical subhalo density profiles that do not have an analytic closed-form expression. 
We hence derive analytic approximations to these stripped subhalo density profiles using a two-step fitting approach, to facilitate a comparison to previously assumed profiles, and to calculate the strength of stream--subhalo encounters. 

First, we find the most consistent numerical profile that matches the \gal{} output. 
\gal{} independently tracks subhalo tidal heating and stripping, producing a heated subhalo density profile and a prediction for the bound mass of the subhalo at each snapshot. 
We apply a truncation function $T(r, r_t) = (1 + r^2/r_t^2)^{-1}$ to the numerically tabulated density profile, fitting for the truncation radius $r_t$ that produces an integrated mass equal to the bound mass reported by \gal{}. 
We then fit the circular velocity curve of this truncated numerical density profile with an analytic truncated Navarro-Frenk-White profile (TNFW; \citealt{Navarro1997, Baltz2009}), keeping $r_s$ fixed to its infall value and fitting for the truncation radius $r_t$ that best matches the circular velocity curve of the subhalo:
\begin{equation}\label{eq:tnfw}
    \rho_\mathrm{TNFW}(r) = \frac{\rho_s}{(r/r_s)(1+r/r_s)^2} \frac{r_t^2}{r^2 + r_t^2}
\end{equation}
The choice of a TNFW profile is well-motivated since \gal{} subhalos are initialized with NFW profiles at infall---it will be valuable for future work to examine this assumption in cosmological simulations of subhalos \citep[e.g.,][]{Dacunha2026}. 
The scale density $\rho_s$ is constrained to match the bound mass reported by \gal{}, resulting in an analytic TNFW profile $\rho_\mathrm{TNFW}(r \mid M_\mathrm{bound}, r_s, r_t)$ for each subhalo. 
Since the profile shape evolution over the duration of our simulation is relatively small for most subhalos, this fit is performed for each subhalo at $z = 0$. 
For snapshots at earlier times, the same $r_s$ and $r_t$ values are reused, but the normalization $\rho_s$ is rescaled to match the evolving bound mass reported by \gal{}.

\begin{figure}
    \centering
    \includegraphics[width=\columnwidth]{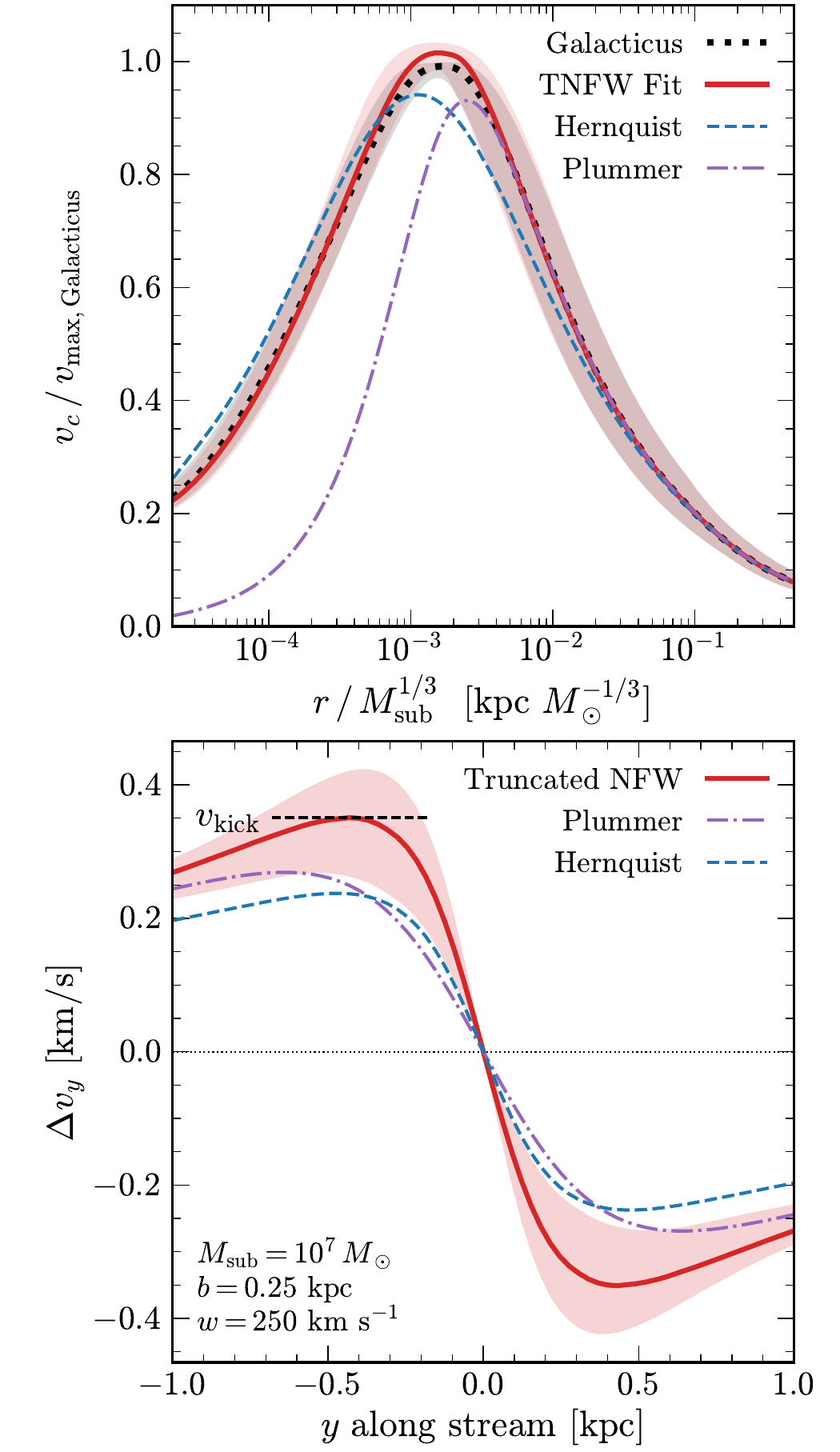}
    \caption{\textbf{The adopted density profile of subhalos in our \gal{} simulation. Top:} The average normalized circular velocity curve is shown for all subhalos within $r_\mathrm{gal} < 100$~kpc, using the truncated numerical profile reported by \gal{}. 
    Our truncated NFW fit is shown in red, which reproduces the numerical curve over all radii. 
    For comparison, Plummer and Hernquist models described by \cite{Erkal2016a}---which encode their own deterministic mass--size relation---are also shown.
    The radial coordinate $r/M_\mathrm{sub}^{1/3}$ is rescaled by the subhalo mass to enable a comparison across all subhalo masses. 
    \textbf{Bottom:} The along-stream velocity perturbation imprinted by a subhalo flyby with the indicated parameters, under the three different profile models. 
    The maximum peak of this curve is adopted as our $\vkick$ statistic to quantify the influence of each impact. 
    }
    \label{fig:subdensity}
\end{figure}

The top panel of Figure~\ref{fig:subdensity} illustrates our fit to the \gal{} subhalo density profile, comparing the average truncated numerical profile for all subhalos within $100$~kpc of our host realizations to the fitted TNFW profile.
The red shaded band shows the $1\sigma$ scatter in the profiles across subhalos. 
We also show corresponding profiles following the prescription described in \cite{Erkal2016a}, which specifies the scale radius for Plummer and \cite{Hernquist1990} profiles fit to subhalos in the Via Lactea II simulation suite \citep{Diemand2008}. 
The bottom panel of Figure~\ref{fig:subdensity} shows the along-stream velocity perturbation imprinted by a subhalo flyby with the indicated parameters, calculated analogously to Equation~2 in \cite{Erkal2015}. 
The peak of this along-stream component is later adopted in this work as the $\vkick$ statistic used to quantify the strength of stream--subhalo encounters (Equation~\ref{eq:vkick}). 
The perturbation profiles imprinted by the Plummer and Hernquist models assumed by \cite{Erkal2016a} are noticeably shallower than those predicted by \gal{} (see also \citealt{Menker2026}), and would predict smaller stream perturbations for close encounters at fixed subhalo mass. 
Two effects contribute here: differences in the shape of the density profiles, and differences in the mass--size relation, with the models assumed by \cite{Erkal2016a} having puffier subhalos than those predicted by \gal{}. 
We therefore proceed with the fitted TNFW profiles as our analytic representation of the \gal{} subhalos, and emphasize that it is important to accurately model subhalo density profiles (and their scatter at a given $M_\mathrm{sub}$) to connect stream perturbations to the underlying subhalo population. 
We investigate the mass--size relation of stream-encountering subhalos further in $\S$\ref{sec:results_subhalos}.

\subsection{The Large Magellanic Cloud}\label{sec:lmc}

The distribution of DM subhalos around the MW is not expected to be spherically symmetric. 
Apart from the influence of the disk, the global potential of the MW is also substantially influenced by the neighboring Large Magellanic Cloud (LMC; \citealt{Garavito-Camargo2019, Garavito-Camargo2021, Erkal2021, Petersen2021, Chandra2025, Brooks2025}). 
The LMC contributes its own population of subhalos, potentially boosting the impact rates of streams in the southern hemisphere \citep{Barry2023, Arora2024}. 

We incorporate a simplified implementation of the LMC's subhalos into our \gal{} simulation by adding the LMC's subhalos ``on-rails'' as a post-processing step onto the \gal{} outputs.
First, we determine the past orbit of the LMC using \texttt{agama}, integrating its current phase-space coordinates back in time in a MW potential that includes the effect of dynamical friction \citep{vanderMarel2002,Pietrzynski2019, Gaia2021}.
This defines the 6D vector of the LMC center-of-mass positions and velocities for all timesteps in our simulation, relative to the MW center-of-mass. 
For each random tree for which we simulate a MW halo, we also simulate in \gal{} an isolated LMC with a host mass of $1.8 \times 10^{11}\,\msun$ \citep{Garavito-Camargo2019, Erkal2019a, Chandra2025}. 
At each simulation timestep, we paint on the LMC subhalos, modifying their positions and velocities to follow the LMC center-of-mass orbit. 
This simplified treatment ignores the gravitational influence of the MW on the LMC's subhalo population, and any induced deformation in the distribution of subhalos. 
However, it captures the first-order influence of the LMC on stream--subhalo interactions, which is the increased flux of subhalos (with large characteristic velocities relative to the MW) in the Galactic southern hemisphere. 
In $\S$\ref{sec:discuss} we comment on ways to improve the fidelity of the LMC's influence when forecasting subhalo populations. 

\subsection{Subhalo Simulation Realizations}

For a given random seed, we run \gal{} with the above described modifications for a $1 \times 10^{12}\,M_\odot$ host MW halo mass \citep[e.g.,][]{Wang2020}. 
Because our \gal{} simulations contain no baryonic physics other than the analytic disk potential described above, we adopt a DM-only cosmology so that each host carries its full virial mass in DM.
Our simulation uses a global time grid from the present day to $-4$~Gyr in the past, in $1$~Myr increments, for a total of $4000$ timesteps. 
Snapshots are produced every $1$~Myr from $T = -4$~Gyr to $T = 0$, recording the evolution of subhalo masses and orbits. 
For computational simplicity, we only consider subhalos that survive (have present-day masses above $M_\mathrm{sub} > \mthresh\,M_\odot$) to $z = 0$.
We run 16 random realizations in \gal{} with different random seeds, storing the full mass evolution of their subhalos. 
To generate additional statistical realizations of our simulation, we also perform 16 rotations for each \gal{} realization. 
Specifically, at $T = -4$~Gyr we rotate the MW and LMC subhalos around their respective host centers of mass by a randomly drawn angle, and use these rotated trajectories at each timestep (rotating the orbits themselves, not re-integrating them). 
The MW host potential with a disk is axisymmetric, not spherically symmetric, but we expect this to have a minor influence on the anisotropy of surviving $10^6\,\msun{}$ subhalos \citep{Green2022}. 
Under the assumption that the subhalo orbits are reasonably isotropic around each host, the rotations produce additional statistical realizations of the same underlying subhalo orbital distributions, for a total of 256 simulation realizations. 
We emphasize that these realizations are not entirely independent---although the rotated realizations reduce the Poisson uncertainty on expected stream interactions, the host--host scatter is captured by the 16 independent \gal{} trees. 

\section{Modeling Stellar Streams}\label{sec:streams}

\subsection{Selecting Candidate Streams}

In large part due to the advent of \textit{Gaia} data, over a hundred candidate stellar streams are known in and around the MW \citep{Helmi2020, Mateu2022, Bonaca2025}. 
A compendium of \textit{Gaia} members in all known streams was compiled by \cite{Bonaca2025}. 
We apply the following selection cuts to select streams that would be suitable detectors of interactions with DM subhalos, where $N_\mathrm{Gaia}$ is the number of \textit{Gaia} members in \cite{Bonaca2025}, $\ell$ is the physical length of the stream reported by \cite{Bonaca2025}, $w$ is the angular width, and $r_\mathrm{peri}$ is the orbital pericenter:

\begin{align}
    N_\mathrm{Gaia} &> 20 \label{eq:ngaia} \\
    \ell &> 3.0~\mathrm{kpc} \label{eq:length} \\
    w &< 1.0~\mathrm{deg} \label{eq:width} \\
    r_\mathrm{peri} &\geq 3.0~\mathrm{kpc} \label{eq:rperi}
\end{align}

The first three cuts select streams that are well-sampled, long, and thin. 
Streams shorter than $\lesssim 3$~kpc have small cross-sections to record interactions with DM subhalos, and streams wider than $1^\circ$ typically originate from dwarf galaxy progenitors rather than star clusters. 
Note that our cuts are applied on the \cite{Bonaca2025} catalog, so some of our final stream models have physical lengths slightly below $3$~kpc.  
The final cut removes streams that pass too close to the inner regions of the MW. 
Baryonic matter in these inner regions---the stellar bar, spiral arms, and giant molecular clouds---is expected to imprint its own signature on streams, complicating the measurement of DM substructure \citep[e.g.,][]{HATTORI2016, Arora2026}.

\begin{figure}
    \centering
    \includegraphics[width=\linewidth]{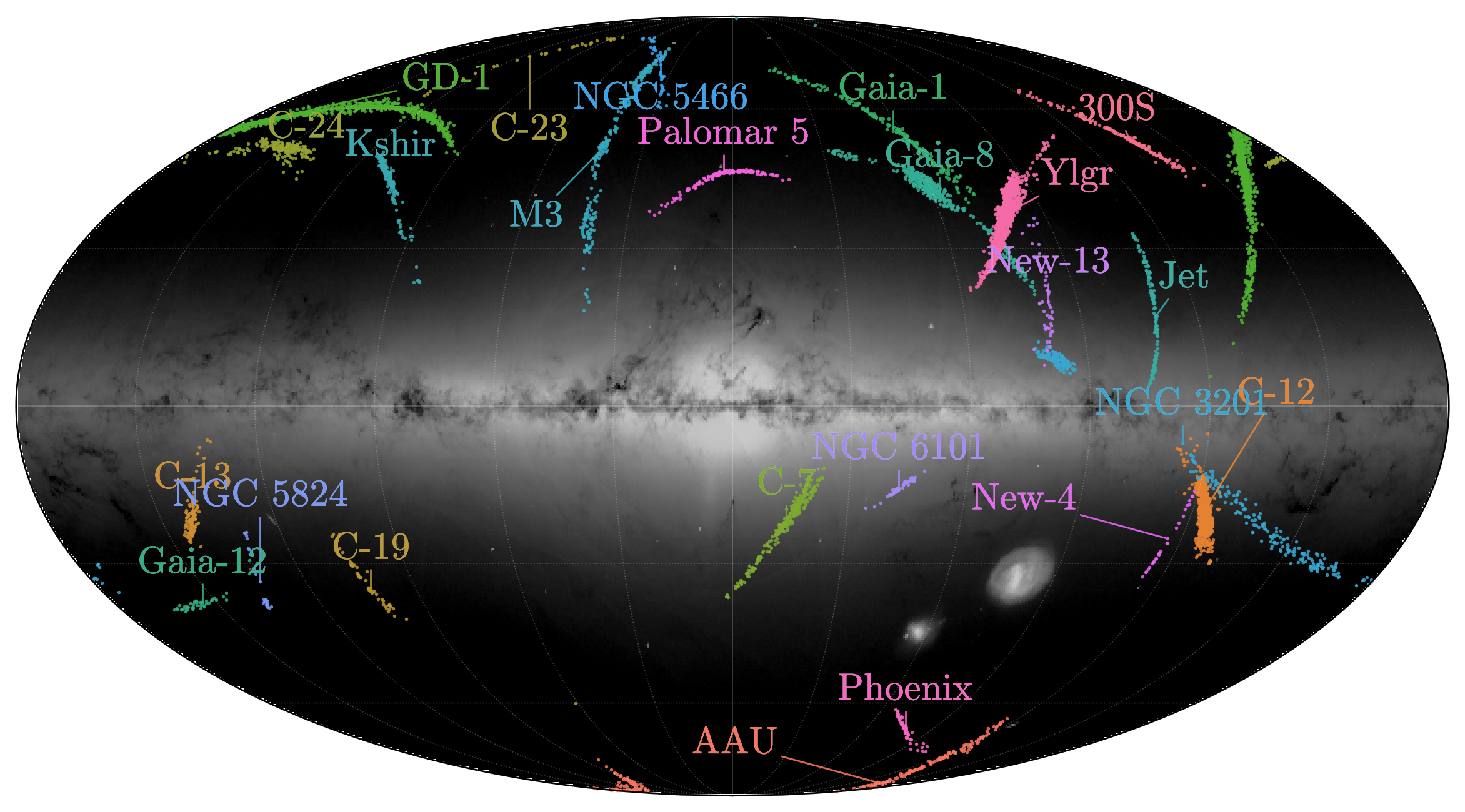}
    \caption{\textbf{On-sky distribution of \textit{Gaia} stars in the globular cluster streams considered in this work.} The stellar data are compiled by \cite{Bonaca2025}, and the distribution of all \textit{Gaia} sources is shown in the background for reference.}
    \label{fig:skymap}
\end{figure}

We also manually exclude some streams that pass these cuts, either because their \textit{Gaia} datasets seem unreliable, their data are too sparse to obtain a reliable progenitor orbital fit (e.g., Gaia-6 and C-20), or because they have been identified as dwarf galaxy streams in the literature (e.g., Orphan-Chenab, Jhelum). 
We retain the 300S stream despite its classification as a dwarf galaxy stream in \cite{Bonaca2025}, since high-resolution spectroscopy finds no significant metallicity dispersion in its member stars \citep{Usman2024}. 
The kinematics of streams with dwarf galaxy progenitors are expected to be substantially influenced by the DM halos of their progenitor galaxies, complicating their use as detectors of the MW's own DM substructure \citep[e.g.,][]{Li2022b}. 
Future work could extend our results to long dwarf-progenitor streams like Orphan-Chenab and Jhelum \citep[e.g.,][]{Koposov2019, Bonaca2019a}, and also streams beyond the primary \cite{Bonaca2025} catalog \citep[e.g.,][]{Grillmair2009, Aganze2025}.
Our selections leave us with $\nstream$ long, cold, and distant globular cluster streams for our analysis.
Their on-sky distribution is shown in Figure~\ref{fig:skymap}. 

\subsection{Simulating the MW's Stellar Streams}\label{sec:streamfit}

In order to generate accurate predictions about the interactions between these streams and DM subhalos, we must develop a model for the evolution of each stream over the duration of our simulation. 
Past works have used various frameworks for computational simplicity, for example assuming that streams follow great circle orbits, or have lengths linearly increasing with time \citep{Erkal2016a, Nibauer2025, Menker2026}. 
In this work, we directly model the evolution of streams through particle-spray simulations, and consequently determine the initial conditions and progenitor properties of each stream \citep{Gibbons2014, Fardal2015, Chen2025}. 

The host potential is defined in \texttt{agama} \citep{Vasiliev2019} including the MW and the LMC, following the orbit integration of the LMC described in $\S$\ref{sec:lmc}. 
The inclusion of the LMC in the host potential can substantially alter the inferred progenitor initial conditions, particularly for streams that pass close to the LMC \citep[e.g.,][]{Shipp2021, Koposov2023, Brooks2024}. 
Although defined differently, the \texttt{agama} host potential used to generate streams is broadly consistent with the \gal{} potential in terms of the components and circular velocity profile. 
The potential components are summarized in Table~\ref{tab:potential}. 

\begin{deluxetable}{lccc}
\tabletypesize{\footnotesize}
\tablecaption{Gravitational potential components of the MW$+$LMC model.\label{tab:potential}}
\tablehead{
  \colhead{Component} & \colhead{Profile} &
  \colhead{Mass} & \colhead{Scale} \\[-0.5em]
  \colhead{} & \colhead{} & \colhead{($10^{10}\,M_\odot$)} & \colhead{(kpc)}
}
\startdata
\multicolumn{4}{l}{\underline{Milky Way}} \\
Stellar disk & $3\times$MN\tablenotemark{a} & $4.77$  & $2.6,\ 0.3$\tablenotemark{a} \\
Bulge        & Hernquist                    & $0.50$  & $1.0$   \\
Nucleus      & Hernquist                    & $0.18$ & $0.07$ \\
DM halo      & TNFW\tablenotemark{b}        & $94$    & $14$    \\[0.3em]
\multicolumn{4}{l}{\underline{LMC}} \\
DM halo      & TNFW\tablenotemark{b}        & $18$    & $12.1$  \\
\enddata
\tablenotetext{a}{\cite{Smith2015} three-component Miyamoto--Nagai (MN) approximation to an exponential disk; scales are the radial and vertical lengths $(r,h)$.}
\tablenotetext{b}{Truncated NFW: \cite{Vasiliev2021a} spheroid with exponential truncation at $r = 300$~kpc for the MW, and $r = 121$~kpc for the LMC.}
\end{deluxetable}

In order to generate a mock stream that matches a given known stream's members, we must determine the structural and orbital properties of the stream's progenitor. 
Our catalog of stream members is drawn from \cite{Bonaca2025}, who provide a compilation of \textit{Gaia} members for all streams considered in our analysis. 
Although these catalogs are not complete and somewhat heterogeneous, they suffice for our present analysis. 
The key unknown variables are the progenitor's $6$D phase-space position $q_\mathrm{prog} \equiv [x,y,z,v_x, v_y, v_z]$, its total stellar mass, and the dynamical age (time since disruption began) of the stream. 
We implement a three-stage fitting procedure to determine these variables, described in detail in the rest of this section. 
Briefly, we first fit the positions, proper motions, parallaxes and (where available) radial velocities of observed stream member stars to fit for $q_\mathrm{prog}$ using a simplified smooth orbital model of the stream. 
Then, we determine the stellar mass of the progenitor $M_\mathrm{prog}$ using isochrones and the \textit{Gaia} members from \cite{Bonaca2025}. 
Finally, we determine the dynamical age by running full particle-spray simulations \citep{Fardal2015, Chen2025} to select the age that best reproduces the observed length of the stream.

\begin{figure}
    \centering
    \includegraphics[width=\columnwidth]{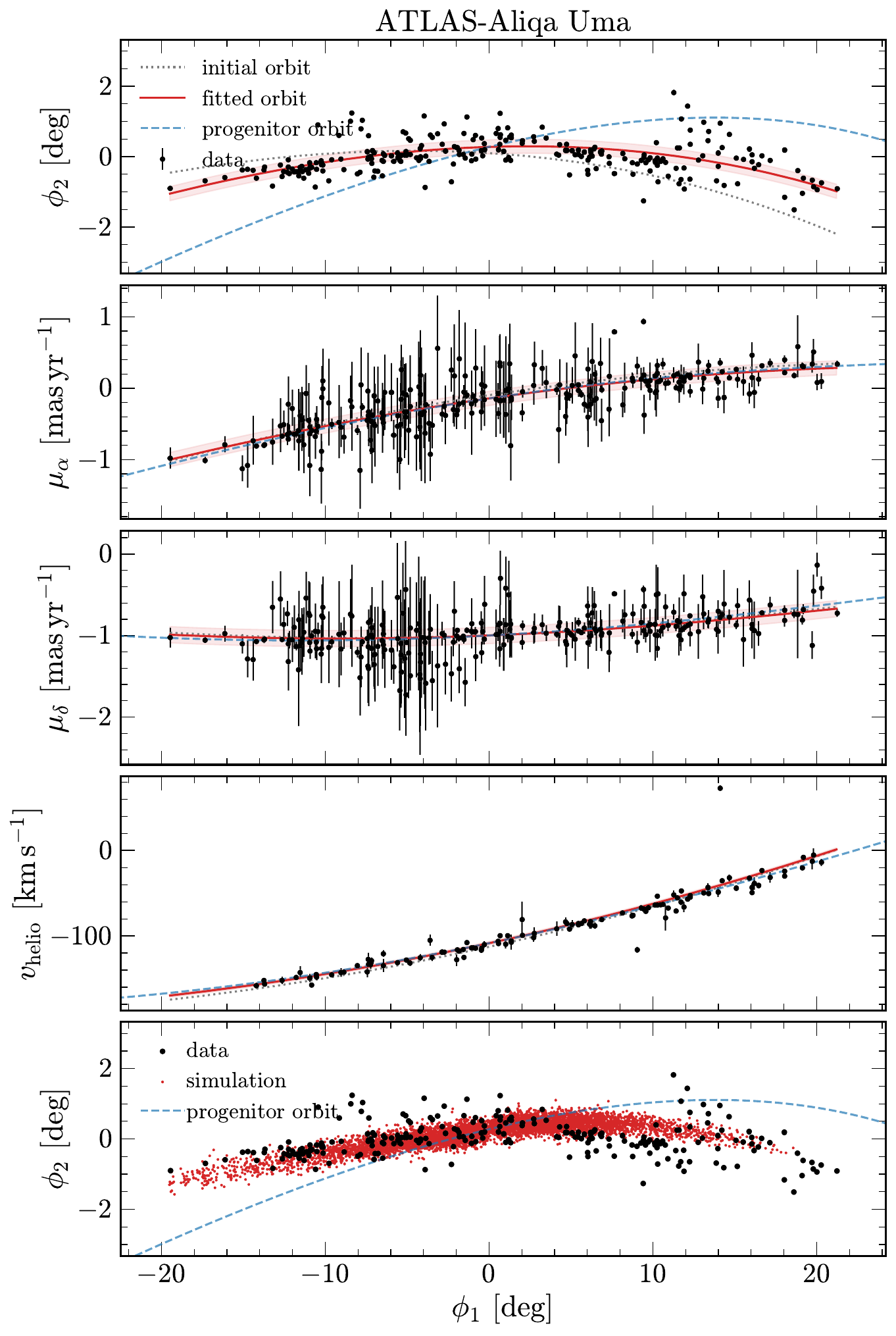}
    \caption{\textbf{Fitting the observed properties of stream members to determine progenitor properties.} The AAU stream is shown here as an example, illustrating the data and fit (from top to bottom) to the on-sky track, proper motions, and radial velocities. 
    The resulting particle-spray streams match the structure, length, and kinematics of each MW stream at the present day (bottom panel).}
    \label{fig:streamfit}
\end{figure}

For a given progenitor phase-space coordinate $q_\mathrm{prog}$, we integrate the progenitor back in time for $4$~Gyr in our time-varying potential that includes a rigid LMC. 
We then select a long segment of the orbit $\approx 4$~Gyr ago and treat these orbit points as particles in a simplified orbital representation of the stream. 
Although this approach does not model the energy offset of stream particles as in a true ``streakline'' model \citep{Kupper2012}, the influence on the mean stream track is minor and hence sufficient for our purposes, producing particle-spray streams that well-match the observed data (see Figure~\ref{fig:streamfit}). 
Each realization of this smooth stream orbital model is evolved forward in the time-varying potential, capturing the response of the tidal tails to the LMC. 
At $T = 0$, the stream model is transformed to angular coordinates on-sky and its phase-space coordinates are interpolated in $\phi_1$ to produce a smooth stream model. 
This model is compared to the observed \textit{Gaia} members with a $\chi^2$ statistic measured (and summed) over the sky position $\phi_2$, equatorial proper motions $\mu_\alpha$ and $\mu_\delta$, parallax $\varpi$, and (wherever available) radial velocity $v_\mathrm{helio}$. 
In addition to the observational uncertainties for each star, we add in quadrature $0.2^\circ$, $0.1$~mas/yr, and $2$\,\kms{} to the position, proper motion, and radial velocity errors when calculating the $\chi^2$ statistic. 
This smooths the objective function, preventing a small number of bright stars from dominating the fit. 
The objective function is minimized using the global simulated annealing algorithm implemented in \texttt{scipy.optimize} \citep{Virtanen2020}. 
In practice, we use the median phase-space position of observed stream stars with $|\phi_1| < 10^\circ$ as an initial starting guess $q_\mathrm{prog,0}$, and fit for $\Delta q_\mathrm{prog}$ deviations from this estimate.

Figure~\ref{fig:streamfit} illustrates the best-fit progenitor model to the ATLAS--Aliqa Uma (AAU; \citealt{Li2021aau}) stream.
In addition to the best-fit model shown in red, we show an initial guess that uses the median phase-space coordinates of the central observed stream stars as the progenitor's phase-space coordinate.
This guessed orbit is clearly a poor match to the stream data, particularly on-sky and in radial velocities. 
The fitted orbit well-represents the mean trend of the observed stream members in all coordinates. 
The figure also shows the progenitor's own orbit as a blue dashed curve. 
For streams like AAU, where the stream stars are significantly affected by the gravitational influence of the LMC, the fitted stream track differs substantially from the progenitor orbit. 
This highlights the value of integrating a full orbital segment of the stream when performing the progenitor orbit fit, allowing the simulated tidal tails to self-consistently react to the time-varying MW potential. 

To determine the approximate stellar mass of the stream progenitor, we assume that the observed streams are fully disrupted, and that the \textit{Gaia} catalogs are complete down to the limiting magnitude of the $99^\mathrm{th}$-percentile faintest star in the sample. 
We then use a 13~Gyr MIST isochrone \citep{Choi2016} and the \cite{Kroupa2001} IMF to estimate what fraction of the stellar population is represented by the \textit{Gaia} members, and consequently calculate the total stellar mass contained in the stream. 
For the isochrone metallicity, we utilize literature metallicity values for each stream where available, and otherwise default to [Fe/H]~$=-2.0$. 
Although this method is approximate---for example, some streams like Palomar~5 have known intact progenitors---our estimates are overall unbiased compared to those in \cite{Bonaca2025}, with a scatter of $\approx 30\%$.

With the progenitor orbit and stellar mass determined, the remaining unknown is the dynamical age of the stream: the time since the tidal disruption of the observed stream extent began. 
For a given dynamical age $\tau_\mathrm{stream}$, the progenitor is integrated back to that time $T = -\tau_\mathrm{stream}$, and then a particle spray simulation is run forward in time. 
We use the \texttt{agama} implementation of the \cite{Chen2025} particle-spray algorithm, which is tuned to match predictions from $N$-body simulations of cluster disruption. 
The angular length (specifically, the range in $\phi_1$ that contains 99\% of particles) of the simulated stream at $T = 0$ is compared to the observed length of the stream based on its \textit{Gaia} members. 
The one-dimensional root-finding algorithm \texttt{scipy.optimize.brentq} is used to determine the best-fitting $\tau_\mathrm{stream}$ that reproduces the length of the observed stream. 
The \cite{Chen2025} algorithm improves upon past algorithms \citep[e.g.,][]{Fardal2015} by modeling the position-dependent dispersion of stars leaving the cluster as it orbits the galaxy, tuned to $N$-body simulations. 
In particular, the \cite{Chen2025} algorithm more accurately partitions the velocity dispersion of stars along the stream's orbit, resulting in a more gradual growth of the tidal tails. 
A consequence of this is that stream ages inferred with the \cite{Chen2025} algorithm are $\approx 50\%$ older than those inferred with the \cite{Fardal2015} algorithm.

The bottom panel of Figure~\ref{fig:streamfit} shows the final simulated stellar stream at $T = 0$ once $q_\mathrm{prog}$, $M_\mathrm{prog}$, and $\tau_\mathrm{stream}$ have been determined. 
We emphasize that for the current work, an exact match to the present-day properties of the stream is not required. 
Rather, we aim to realistically capture the dynamical evolution of each stream over the past $\approx 4$~Gyr, in order to accurately estimate the abundance of stream--subhalo interactions. 
The three-stage fitting methodology described here serves as a generic and accurate technique to produce mock models for any stream in a realistic time-varying potential. 
The method can be improved by incorporating more information about the progenitor's location and properties when it is known, and fitting for the progenitor location when it is unknown. 
We also do not explicitly fit any distance data (apart from the parallax, which is minimally informative for most distant streams).
Future work could incorporate stellar distance catalogs of known streams \citep[e.g.,][]{Chandra2026, DOMINGUEZ2026}. 

\begin{figure}
    \centering
    \includegraphics[width=\linewidth]{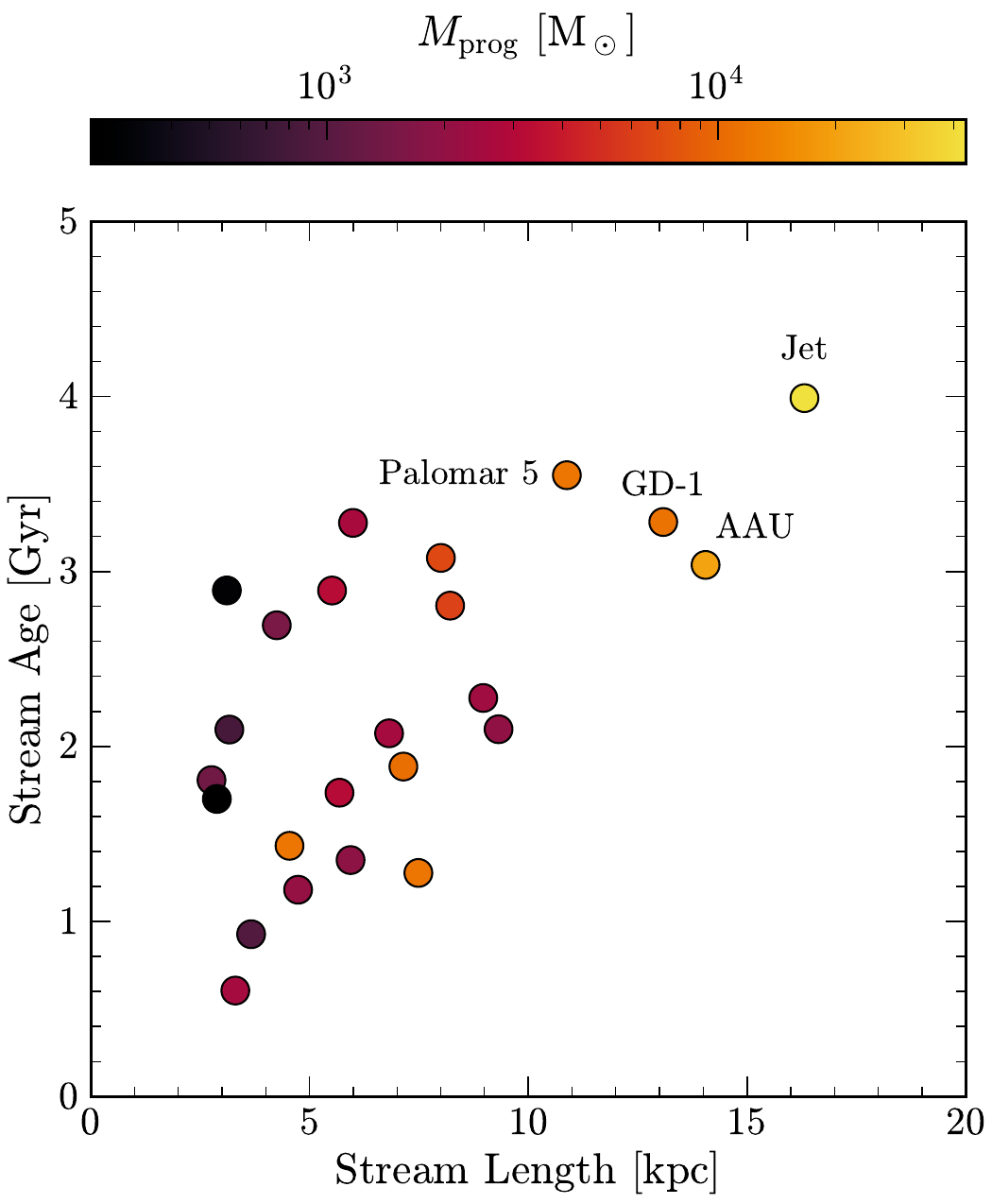}
    \caption{\textbf{Progenitor properties of streams considered in this work, as determined from our fits to the \textit{Gaia} members.}
    The stream age is the dynamical age required to produce the observed extent of the \textit{Gaia} members---the stream progenitors likely began disrupting at even earlier times.}
    \label{fig:streamfit_summary}
\end{figure}

Figure~\ref{fig:streamfit_summary} summarizes the fit results for our sample of streams. 
Physically longer streams generally have older dynamical ages, and more massive progenitors. 
This is the expected outcome of the model, since tidal tails grow over time, and more massive progenitors contain enough stars to source longer-lived streams. 
However, the orbital properties of each stream also play an important role in determining how long they grow after a given dynamical timescale, since stars are stripped at various locations in the Galactic potential. 
Furthermore, the present-day observed length of a stream is strongly dependent on its orbital phase; for streams on eccentric orbits, the projected length is much longer near pericenter than near apocenter. 
The stream modeling described in this section aims to capture these effects directly to derive representative progenitor properties and dynamical ages, which directly influence the accumulation of subhalo perturbations over time. 

\subsection{Quantifying stream--subhalo Impacts}

\begin{figure}
    \centering
    \includegraphics[width=\linewidth]{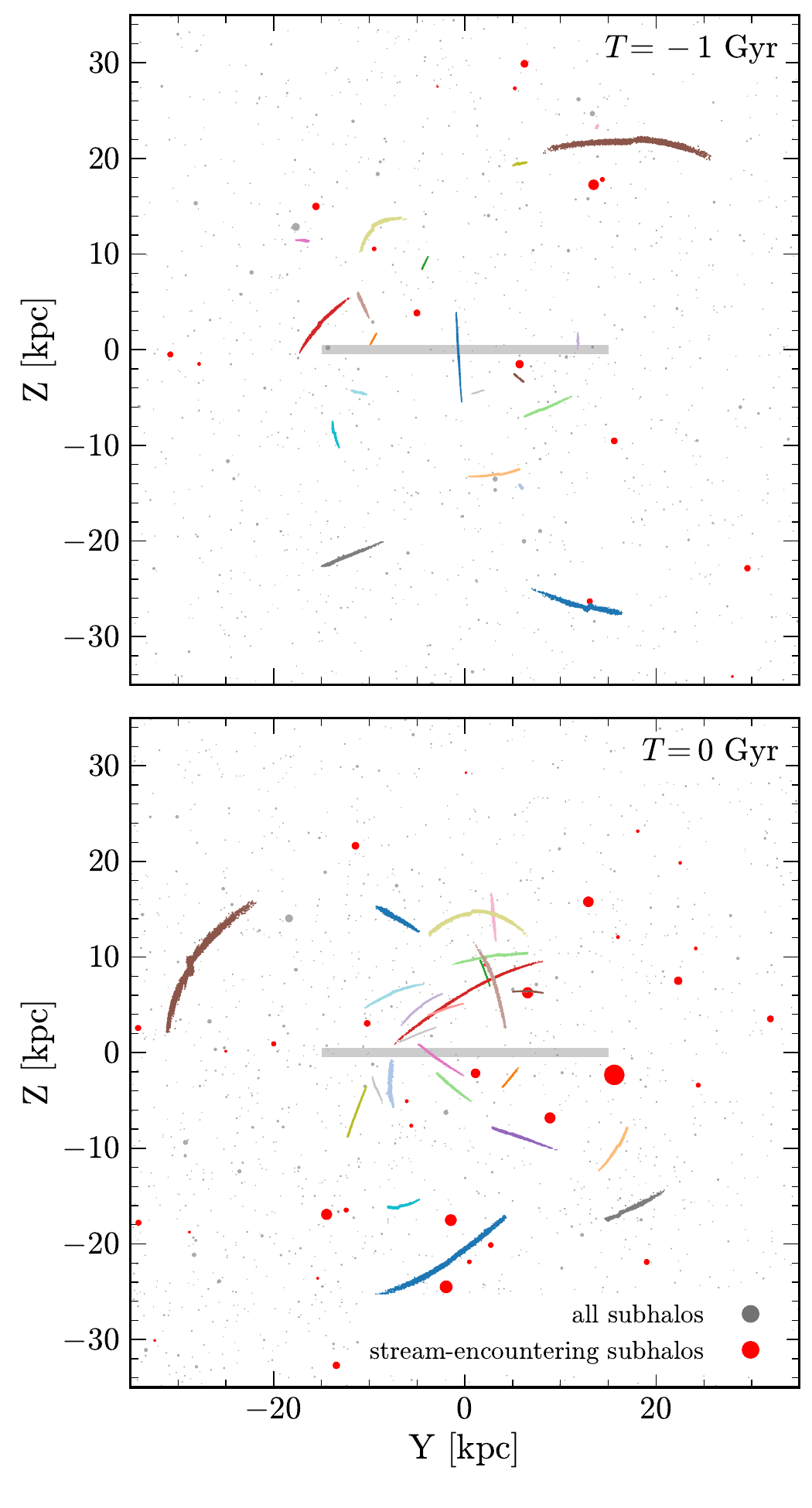}    \caption{\textbf{Simulating interactions between MW streams and \gal{}-predicted subhalos.} Two snapshots from our simulation for a single \gal{} realization are shown, highlighting the growth and evolution of each stream. All $> \mthresh{}\,\msun{}$ subhalos are shown in grey; those that encounter any MW stream and impart a $\vkick > 0.1$\,\kms{} perturbation are highlighted in red. The approximate extent of the MW disk is indicated with a grey bar. An animated version of this figure is available\footnote{\url{https://youtu.be/8lnZPWZCOkM}}.}
    \label{fig:impactmovie}
\end{figure}

We are now equipped with a simulation that models the growth and evolution of the MW's stellar streams from $T = -4$~Gyr to $T = 0$~Gyr on a common time grid as the \gal{} simulations of the MW and LMC's subhalo populations (see Figure~\ref{fig:impactmovie}). 
A direct numerical technique is implemented to quantify the encounters between streams and subhalos. 
At each timestep, for each subhalo, the closest distance to any particle in the stream is recorded. 
A peak-finding algorithm (\texttt{scipy.signal.find\_peaks}) determines all the minima that denote closest-approach encounters.
We pre-select subhalos that have close encounters as those that pass by the stream within $10\times$ the subhalo scale radius at infall \citep{Erkal2016a}. 
This is a relatively generous and complete selection, since we later classify the impacts more stringently based on the velocity kicks they impart. 
% We also only consider encounters whose closest approach ends up within the central $90\%$ of the stream in terms of present-day angular $\phi_1$ extent---this filters out encounters that only influence the very tips of the streams simulated, where there are few observable stars. 
This collection of impulsive encounters forms the database of encounters that are further analyzed in this work. 
In this framework, a single subhalo may have multiple impacts with the same stream if they lie on similar orbits, and a given subhalo may encounter multiple streams. 

With the collection of impulsive stream--subhalo impacts in hand, we can calculate several summary statistics to characterize the impact. 
The most fundamental impact property is the velocity kick $v_\mathrm{kick}$ it imparts on the stream stars near the subhalo, in the impulse approximation.
This is the dominant long-lived perturbation imparted by subhalo flybys, and over long timescales leads to the formation of density substructures like gaps \citep{Yoon2011, Erkal2015, Erkal2016a, Sanders2016}. 
As such, $v_\mathrm{kick}$ serves as a compact metric to describe the amplitude of a subhalo interaction. 

In practice, the subhalo flyby can impart a velocity perturbation along multiple components of the stream members' velocity distributions. 
Past works have focused on the gap-forming influence of subhalo encounters, which is encoded by the along-stream (stream-rotated $y$ axis by convention, with $x$ towards the Galactic center) velocity perturbation \citep[e.g.,][]{Adams2025, Menker2026}. 
We therefore use the amplitude of the differential velocity kick $|\Delta v_y|_\mathrm{max}$ as a compact metric quantifying the strength of the encounter. 
Specifically, we use the maximum value of this kick velocity along the stream as our summary statistic $\vkick \equiv |\Delta v_y|_\mathrm{max}$.

We follow the derivation in \cite{Erkal2015} for velocity perturbations from subhalo flybys in the impulse approximation, replacing their assumed Plummer profiles with the TNFW profiles. 
This results in a numerically calculable form for $\Delta v$ along each velocity component, from which $\vkick \equiv |\Delta v_y|_\mathrm{max}$ is derived for each encounter in our simulation
\begin{equation}\label{eq:vkick}
    v_\mathrm{kick} \equiv  |\Delta v_y|_\mathrm{max} = \frac{G M_\mathrm{sub} w_\perp}{w^2 b} \, \mathcal{F}%(\frac{b}{r_s}, \frac{r_t}{r_s})
\end{equation}
where $M_\mathrm{sub}$ is the subhalo mass at the encounter time, $w$ is the total relative flyby velocity, $w_\perp$ is its component perpendicular to the stream, $b$ is the impact parameter relative to the closest point on the stream, and $\mathcal{F}$ is a dimensionless numerical function that encodes the softening caused by the subhalo's extended mass profile.
Explicitly, 
\begin{equation}\label{eq:F}
    \mathcal{F} = \max_{B \geq b} \left[ \frac{2\,b\,\sqrt{B^2 - b^2}}{B^2} \; \frac{M_\mathrm{2D}(<B)}{M_\mathrm{sub}} \right]
\end{equation}
where $M_\mathrm{2D}(<B)$ is the projected subhalo mass within a cylinder of radius $B$ along the flyby direction; $\mathcal{F} \leq 1$, approaching unity for encounters distant compared to the subhalo's size.
For a Plummer profile of scale radius $a$, this function simplifies to $\mathcal{F} = b/\sqrt{b^2+a^2}$ \citep{Erkal2015}. 
For the TNFW profiles used here, we numerically evaluate $\mathcal{F}$ for each encounter to calculate $\vkick$. 

We perform several cuts to select localized perturbations for which the impulse approximation holds. 
We discard encounters with subhalos more massive than $M_\mathrm{sub} > 10^9\,M_\odot$, which tend to produce global rather than local perturbations. 
Additionally, we filter impacts based on the characteristic size of the impact footprint. 
As the lower panel of Figure~\ref{fig:subdensity} shows, the asymmetric component of the perturbation has a characteristic size along the stream, which depends almost entirely on the encounter geometry. 
The two extrema of $\Delta v_y$ lie at $\pm\Delta y$ from the point of closest approach, with $\Delta y = (w/w_\perp)\sqrt{B_\star^2 - b^2}$ where $B_\star$ is the value of $B$ that maximizes Equation~\ref{eq:F}. 
We use this kick width numerically calculated for the TNFW profile to select encounters that impart local density and velocity perturbations to the stream, retaining encounters with $\Delta y < 5$~kpc. 
This removes distant encounters (predominantly with more massive $> 10^8\,M_\odot$ subhalos) that secularly shift the entire orbit of the progenitor and stream, and retains encounters that impart local differential density and velocity variations on streams.

\begin{figure}
    \centering
    \includegraphics[width=\linewidth]{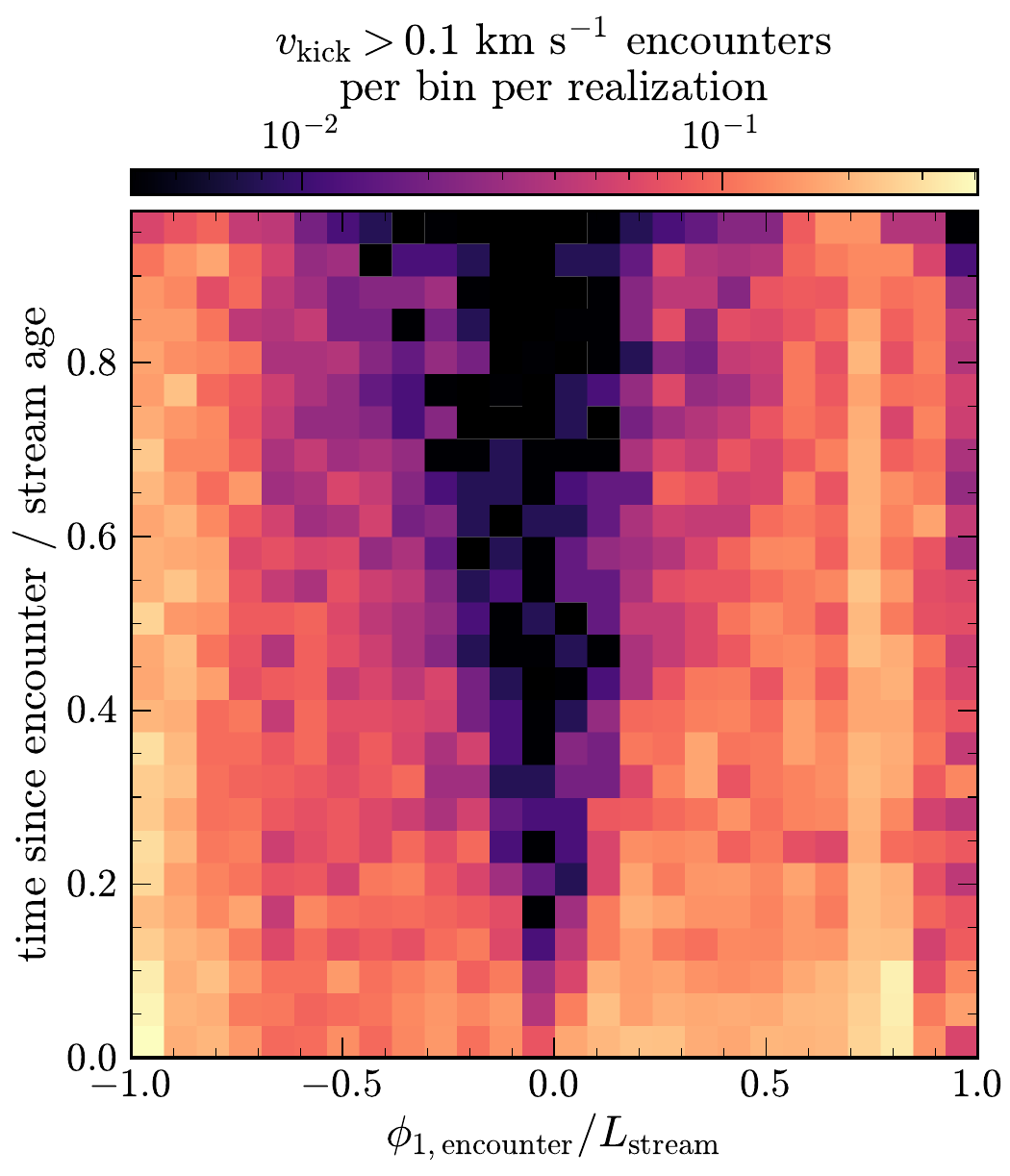}
    \caption{\textbf{When and where do subhalos encounter the MW's streams?} The distribution of $\vkick > 0.1$\,\kms{} encounters is shown as a function of the present-day $\phi_1$ location of the encounter's closest approach (normalized by the length of each tail of the stream) and the time since the encounter normalized by the dynamical age of each stream. 
    }
    \label{fig:phi1_trel}
\end{figure}

Figure~\ref{fig:phi1_trel} summarizes the present-day encounter location along the stream, and time since the subhalo encounter, for our entire simulation. 
This highlights several aspects of the direct simulation, chiefly how different observed locations along the stream are sensitive to different impact timeframes, with the earliest-stripped outer tails naturally being sensitive to the entire history of encounters over the age of the stream. 
The central regions of the stream closest to the progenitor are preferentially sensitive to the most recent encounters.
Our selection criteria include encounters that impart a single velocity perturbation peak on the stream, influencing the morphology and velocities of the outer tails (see Figure~\ref{fig:phi1_trel}).
If only gap-forming encounters are selected that place both velocity perturbation peaks on the stream (see the bottom panel of Figure~\ref{fig:subdensity} and \citealt{Adams2025}), the encounter rates presented here would be adjusted down by roughly a factor of two. 
It will be valuable to incorporate this geometric information into analytic forecasting and inference frameworks, when predicting which regions of a stream are perturbed by a given underlying distribution of subhalos \citep[e.g.,][]{Erkal2016a, Nibauer2026}. 
For example, if only the central region of streams containing $90\%$ of the member stars is reliably observable, the encounter rates presented here would be lowered by $\approx 30\%$. 
Conversely, observational surveys might preferentially target outer stream tails with deep observations to measure subhalo perturbations, complemented by observations of the inner tails to calibrate the intrinsic velocity dispersion profile of the stream.

\section{Results}\label{sec:results}

\subsection{Encounter Rates of MW Streams}

The degree to which a stream has been influenced by dark subhalos can be compactly summarized by the number of impulsive impacts that deliver a velocity kick $v_\mathrm{kick}$ larger than some threshold value. 
We proceed by examining encounters with $v_\mathrm{kick} > 0.1$~\kms{} as those that impart signatures which could be detectable by morphological and kinematic information \citep[e.g.,][]{Erkal2016a, Adams2025, Lu2026}. 
In practice, encounters can be inferred from a variety of measurements including direct RV signatures (Figure~\ref{fig:impact_sim}), power spectra of density and velocity variations, and integrated measurements like velocity dispersions \citep[e.g.,][]{Erkal2017, Bovy2017, Ibata2020, Banik2021streams, Nibauer2026}. 
We discuss the detectability of encounters further in $\S$\ref{sec:discuss}, but leave a full observational forecast to future work. 
The properties and encounter rates of streams considered in this work are summarized in Table~\ref{tab:streams}.

\begin{figure}
    \centering
    \includegraphics[width=\columnwidth]{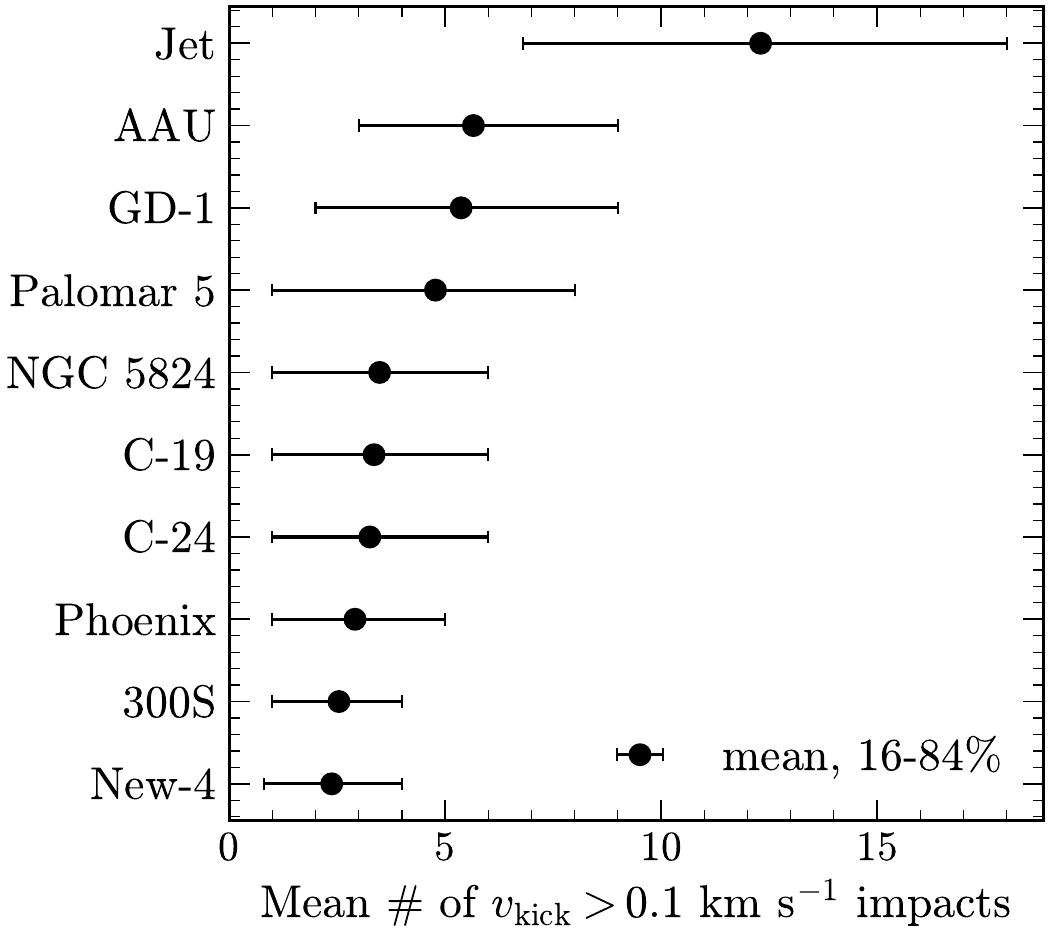}
    \caption{\textbf{The number of CDM subhalos encountered by MW streams over their lifetimes.} Points indicate the mean number of significant $\vkick > 0.1$\,\kms{} subhalo encounters, and error bars indicate the $1\sigma$ percentile range of the across-realization scatter. 
    Streams are arranged by the mean number of such encounters, and only streams with $> 2$ predicted encounters are shown.}
    \label{fig:ndetect}
\end{figure}

\begin{figure}
    \centering
    \includegraphics[width=\columnwidth]{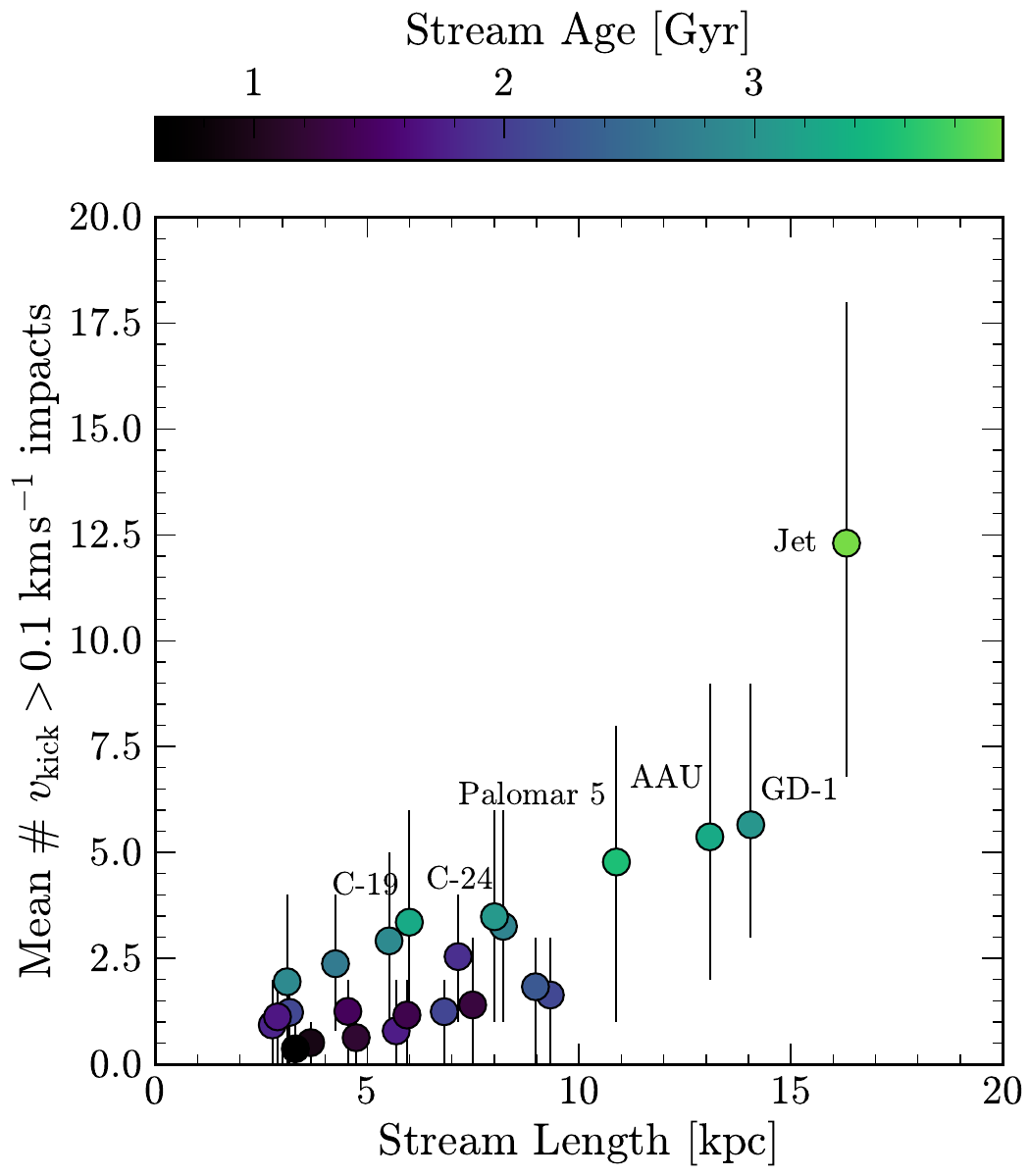}
    \caption{\textbf{The number of stream--subhalo encounters as a function of the stream's physical length and dynamical age}. Physically long and long-lived streams have the largest number of predicted encounters.}
    \label{fig:lengthdetect}
\end{figure}

Figure~\ref{fig:ndetect} summarizes the mean number of $v_\mathrm{kick} > 0.1$~\kms{} encounters experienced by each stream, with the $1\sigma$ scatter across simulated realizations indicated. 
Streams are arranged by the mean number of such encounters across realizations, and only streams with $> 2$ predicted encounters are shown. 
The characteristic scatter of the encounter rates across \gal{} realizations (aggregated across the top 10 streams) is $\approx 40\%$, set by the host-to-host scatter in the number density of subhalos in the inner regions of the galaxy. 
Note that the scatter in the encounter rate predicted for any individual stream has a substantial Poisson component, with the total per-stream scatter rising to $\approx 70\%$. 
Figure~\ref{fig:lengthdetect} shows the number of encounters as a function of the stream length, colored by the stream's dynamical age.
As expected, the length of a stream is the most powerful determinant of the subhalo encounter rate, and long-lived streams accumulate more encounters over their lifetimes.

\begin{figure}
    \centering
    \includegraphics[width=\linewidth]{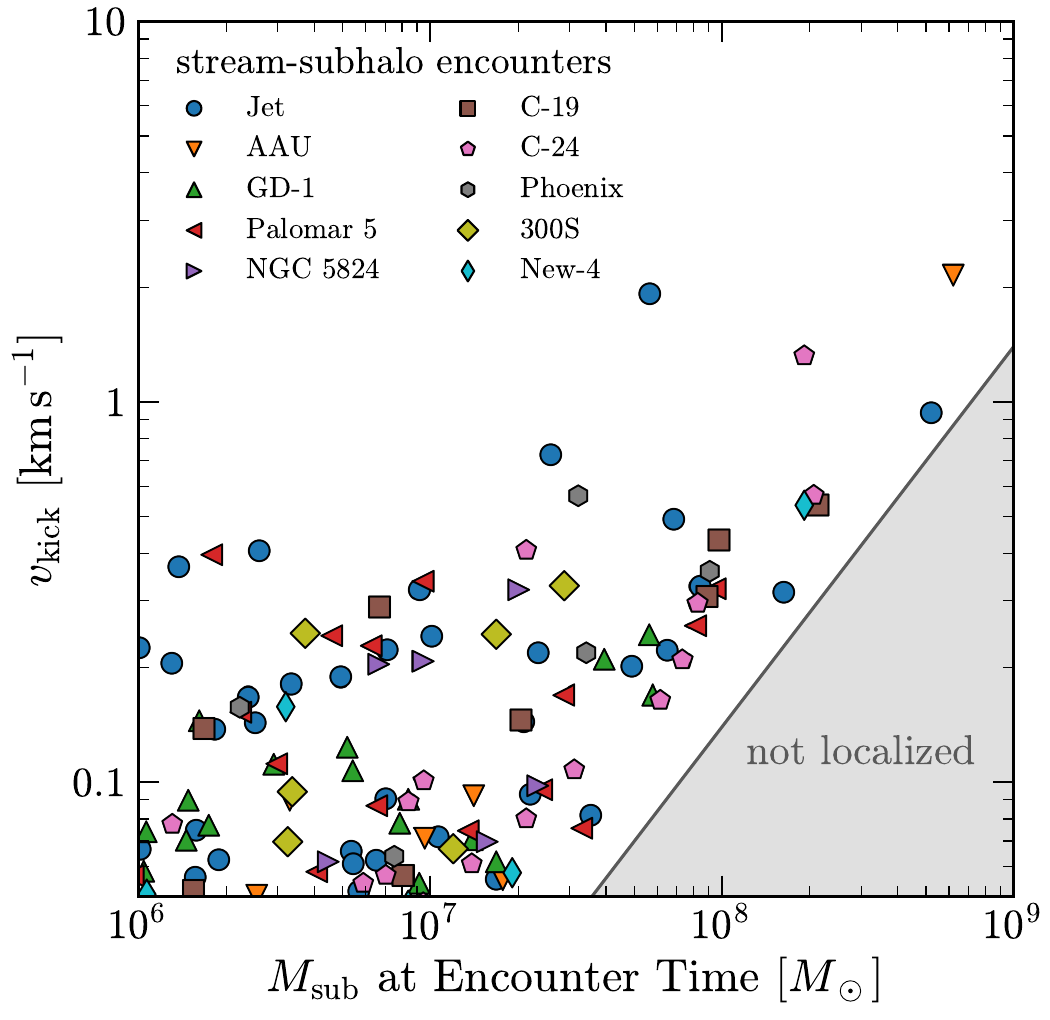}
    \caption{\textbf{Impulsive kick velocities delivered to MW streams as a function of subhalo mass at the time of impact, in a single \gal{} realization}. Each point indicates a single stream--subhalo encounter---at constant mass, larger $\vkick$ corresponds to closer or slower encounters. 
    This panel shows one typical realization out of the 256 realizations computed in this work. 
    }
    \label{fig:vkick_msub}
\end{figure}

Figure~\ref{fig:vkick_msub} summarizes the individual subhalo encounters recorded in a single realization of our simulation. 
The top 10 streams are shown---in terms of their mean number of detectable impacts across realizations---and the realization is chosen to have a typical number of encounters. 
Figure~\ref{fig:vkick_msub} shows the subhalo mass at the time of impact along with the corresponding velocity perturbation $\vkick$ of the impact. 
Note that a single subhalo can appear multiple times in this figure, if it encounters the same stream multiple times, or encounters multiple different streams. 
For any given subhalo at some mass $M_\mathrm{sub}$, Equation~\ref{eq:vkick} predicts a maximum possible kick velocity. 
This explains the approximately diagonal upper envelope in Figure~\ref{fig:vkick_msub}---at each $M_\mathrm{sub}$, the maximum $v_\mathrm{kick}$ observed corresponds to the closest and/or slowest encounters.
The lack of encounters in the lower right of Figure~\ref{fig:vkick_msub} is due to the selection of impulsive encounters that impart a localized differential velocity perturbation to the stream, i.e. which have a characteristic perturbation size $\Delta y < 5$~kpc. 
Encounters in that regime of the panel correspond to distant encounters with more massive subhalos, which may impart a global perturbation to the stream. 

Figure~\ref{fig:vkick_msub} emphasizes that in any given realization, even the most promising MW streams are only expected to have $\mathcal{O}(1)$ strong $v_\mathrm{kick} > 1$\,\kms{} subhalo encounters that might lead to significant morphological features and large velocity signatures detectable by low-resolution spectroscopic surveys \citep[see also][]{Yoon2011, Erkal2016a, Bovy2017}. 
These strong encounters are almost entirely from subhalos with masses greater than $\gtrsim 10^7\,M_\odot$ at the encounter time. 
Reliably detecting subhalo masses $\lesssim 10^7\,M_\odot$ requires a sensitivity to sub-\kms{} $\vkick$ encounters.
This figure also highlights the utility of surveying multiple streams in concert, since the specific encounter history of an individual stream can vary substantially due to Poisson fluctuations and the underlying host-to-host scatter in the number of subhalos predicted by \gal{}. 
Therefore, combining encounter constraints from multiple streams can pin down the amplitude of the MW's own subhalo population down to $M_\mathrm{sub} \gtrsim 10^6\,M_\odot$, if observations are sensitive to $\vkick \lesssim 1$\,\kms{} encounters. 

Several existing techniques model subhalo encounters as independent events with some Poisson rate that depends on the ensemble properties of the subhalo population \citep[e.g.,][]{Erkal2015, Nibauer2025}.
The direct integration structure of our simulation (Figure~\ref{fig:schematic}) enables tracking the encounter characteristics of individual subhalos over time. 
For example, how often does the same subhalo encounter a given stream multiple times? 
Averaged across realizations, between $0$--$20\%$ of $\vkick > 0.1$\,\kms{} perturbations experienced by a given stream are caused by repeat encounters with the same subhalo. 
The prevalence of repeat encounters is almost entirely predicted by the number of radial orbital periods that have elapsed over the stream's dynamical age, since this increases the probability of re-encountering a subhalo with a fortuitously similar orbital configuration as the stream. 
Palomar~5 has the highest proportion of repeat encounters ($\approx 20\%$), due to its relatively short orbital period and long dynamical age. 
We note that our numerical selection for encounters as distinct peaks in stream--subhalo separations could select against subhalos on very similar orbits as streams, which would almost continuously encounter and disrupt a given stream. 

The rate of a single subhalo encountering multiple streams is even larger. 
Of the $\approx 45$ $\vkick > 0.1$\,\kms{} significant encounters with the top 10 MW streams, $25\%$ are from subhalos that significantly encounter at least one other stream on that list. 
Subhalos that impact multiple streams are preferentially on the more massive end, since they have correspondingly larger spheres of influence. 
The upshot of these calculations is that the $\approx 45$ $\vkick > 0.1$\,\kms{} perturbations are induced by $37$ unique subhalos on average. 
Although these measurements require further investigation in more realistic simulations---for example in which the subhalo populations self-consistently react to the infalling LMC---they suggest that it may be useful to incorporate the correlation structure of repeat subhalo encounters into analytic stream--subhalo encounter calculations, and even into inference pipelines. 
A second-order effect (captured by our framework here) is that subsequent encounters are typically at lower subhalo mass, since the subhalos continue to lose mass as they orbit around the MW.  

Four streams stand out as the most promising dark matter detectors: Jet, AAU, GD-1, and Palomar~5 \citep{Rockosi2002, Grillmair2006, Jethwa2018jet, Li2021aau}. 
This is unsurprising, since these are among the longest and longest-lived of all the streams we consider here (Figure~\ref{fig:streamfit_summary}). 
We predict that the Jet stream is a particularly promising target, due in large part to the old inferred age of its visible tidal tails. 
Jet is the most distant stream in our sample, and its large Galactocentric radius results in a more gradual growth of its tidal tails compared to more nearby streams. 
Furthermore, photometric observations of Jet already indicate the presence of morphological variations like gaps, spurs, and fanned tails \citep{Ferguson2022, Do2026}. 

\begin{figure}
    \centering
    \includegraphics[width=\linewidth]{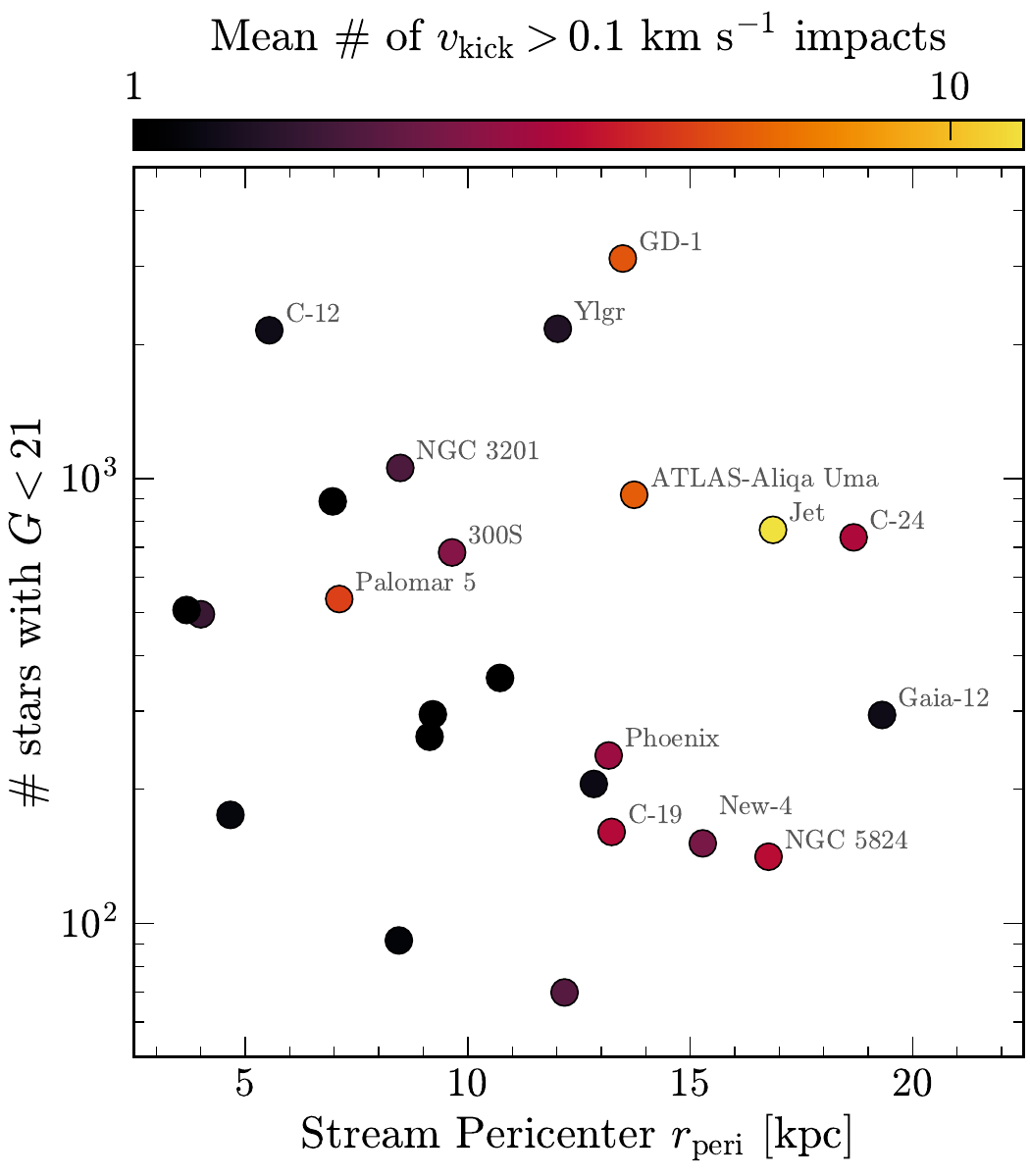}
    \caption{\textbf{The observability and orbits of MW streams.} 
    The number of stars down to the $G < 21$ \textit{Gaia} limit is predicted based on the \cite{Bonaca2025} membership catalogs. 
    Points are colored by the mean number of significant subhalo encounters predicted by our simulation. 
    Streams towards the upper-right are favorable: more observable members and fewer baryonic perturbations from the MW's disk. 
    }
    \label{fig:observability}
\end{figure}

Although we leave a full observational forecast of encounter detectability to future works, we summarize two key observational properties of the streams considered here in Figure~\ref{fig:observability}, colored by the expected subhalo encounter rate. 
The orbital pericenter of the stream $r_\mathrm{peri}$ governs how close a stream gets to the MW disk, exposing it to baryonic perturbations that could imprint their own structure on stream densities and velocities \citep[e.g.,][]{Amorisco2016, HATTORI2016, Arora2026}. 
The number of $G < 21$ stars indicates how many stream stars could be efficiently targeted down to the \textit{Gaia} limit by upcoming spectroscopic surveys like the Via Survey \citep{ViaCollaboration2026} and Spec-S5 \citep{Besuner2025}. 
These numbers are based on our progenitor fit calculations in $\S$\ref{sec:streamfit}, and are dependent on the completeness and purity of the membership catalogs adopted by \cite{Bonaca2025}. 
Streams to the upper-right of this plot are observationally favorable---minimally perturbed by the baryonic influence of the disk, with ample spectroscopically measurable members. 
Several streams with large predicted subhalo encounter rates lie in the favorable region of parameter space: Jet, GD-1, AAU, and C-24. 
Palomar~5 remains a promising target due to its large encounter rate, but may require more careful modeling due to its small pericenter. 
C-19 is relatively sparse down to $G < 21$, and furthermore has an exceptionally low metallicity, which will inflate the RV uncertainties of individual stars \citep{Martin2022}. 

\begin{deluxetable*}{lcccccccccc}[ht]
\tablewidth{0pt}
\tabletypesize{\footnotesize}
\tablecaption{Streams considered in this work.\label{tab:streams}}
\tablehead{\colhead{Stream} & \colhead{$d_\odot$} & \colhead{$M_{\rm prog}$} & \colhead{$\tau_{\rm stream}$} & \colhead{$r_{\rm peri}$} & \colhead{$r_{\rm apo}$} & \colhead{$L$} & \colhead{$L${}} & \colhead{$N_{\rm Gaia}$} & \colhead{$N_{G<21}$} & \colhead{$N_{\rm enc}$}\\ \colhead{} & \colhead{kpc} & \colhead{$10^3\,M_\odot$} & \colhead{Gyr} & \colhead{kpc} & \colhead{kpc} & \colhead{deg} & \colhead{kpc} & \colhead{} & \colhead{} & \colhead{$v_{\rm kick}>0.1$ km\,s$^{-1}$}}
\startdata
Jet & 31.7 & 42.9 & 4.0 & 17 & 37 & 29 & 16.3 & 144 & 800 & $12.3^{+5.7}_{-5.5}$ \\
ATLAS-Aliqa Uma & 21.6 & 20.2 & 3.0 & 14 & 41 & 37 & 14.1 & 245 & 900 & $5.7^{+3.3}_{-2.7}$ \\
GD-1 & 7.6 & 11.2 & 3.3 & 13 & 22 & 98 & 13.1 & 1478 & 3100 & $5.4^{+3.6}_{-3.4}$ \\
Palomar 5 & 20.6 & 11.6 & 3.6 & 7 & 16 & 30 & 10.9 & 129 & 500 & $4.8^{+3.2}_{-3.8}$ \\
NGC 5824 & 28.5 & 6.8 & 3.1 & 17 & 46 & 16 & 8.0 & 27 & 100 & $3.5^{+2.5}_{-2.5}$ \\
C-19 & 19.2 & 2.6 & 3.3 & 13 & 22 & 18 & 6.0 & 45 & 200 & $3.4^{+2.6}_{-2.4}$ \\
C-24 & 13.2 & 6.3 & 2.8 & 19 & 24 & 36 & 8.2 & 244 & 700 & $3.3^{+2.7}_{-2.3}$ \\
Phoenix & 17.5 & 3.2 & 2.9 & 13 & 18 & 18 & 5.5 & 60 & 200 & $2.9^{+2.1}_{-1.9}$ \\
300S & 17.8 & 10.8 & 1.9 & 10 & 67 & 23 & 7.1 & 141 & 700 & $2.5^{+1.5}_{-1.5}$ \\
New-4 & 14.5 & 1.5 & 2.7 & 15 & 22 & 17 & 4.3 & 29 & 200 & $2.4^{+1.6}_{-1.6}$ \\
C-23 & 8.8 & 0.3 & 2.9 & 12 & 18 & 20 & 3.1 & 29 & 100 & $1.9^{+2.1}_{-1.9}$ \\
NGC 3201 & 4.7 & 2.4 & 2.3 & 8 & 29 & 110 & 9.0 & 607 & 1100 & $1.8^{+1.2}_{-1.8}$ \\
M3 & 8.0 & 2.0 & 2.1 & 4 & 20 & 67 & 9.3 & 237 & 500 & $1.6^{+1.4}_{-1.6}$ \\
Ylgr & 10.1 & 11.8 & 1.3 & 12 & 28 & 42 & 7.5 & 919 & 2200 & $1.4^{+1.6}_{-1.4}$ \\
C-12 & 10.0 & 11.7 & 1.4 & 6 & 18 & 26 & 4.5 & 929 & 2200 & $1.2^{+0.8}_{-1.2}$ \\
Gaia-12 & 13.5 & 2.5 & 2.1 & 19 & 78 & 29 & 6.8 & 46 & 300 & $1.2^{+0.8}_{-1.2}$ \\
C-13 & 7.9 & 0.8 & 2.1 & 13 & 15 & 23 & 3.2 & 100 & 200 & $1.2^{+0.8}_{-1.2}$ \\
NGC 5466 & 15.1 & 1.9 & 1.4 & 5 & 51 & 23 & 5.9 & 43 & 200 & $1.2^{+0.8}_{-1.2}$ \\
New-13 & 5.8 & 0.2 & 1.7 & 8 & 35 & 28 & 2.9 & 47 & 100 & $1.1^{+0.9}_{-1.1}$ \\
Kshir & 9.2 & 1.4 & 1.8 & 9 & 13 & 17 & 2.8 & 120 & 300 & $0.9^{+1.1}_{-0.9}$ \\
Gaia-8 & 7.2 & 3.1 & 1.7 & 7 & 13 & 45 & 5.7 & 453 & 900 & $0.8^{+1.2}_{-0.8}$ \\
C-7 & 8.2 & 2.1 & 1.2 & 4 & 96 & 33 & 4.7 & 239 & 500 & $0.6^{+0.4}_{-0.6}$ \\
Gaia-1 & 5.4 & 0.9 & 0.9 & 11 & 70 & 39 & 3.7 & 200 & 400 & $0.5^{+0.5}_{-0.5}$ \\
NGC 6101 & 14.2 & 2.6 & 0.6 & 9 & 44 & 13 & 3.3 & 94 & 300 & $<0.5$
\enddata
\tablecomments{$d_\odot$, $M_{\rm prog}$ and $\tau_{\rm stream}$ are the fitted present-day heliocentric distance, stellar mass and dynamical age of the stream, the time over which its observed extent was stripped from the progenitor (\S\ref{sec:streamfit}); $r_{\rm peri}$ and $r_{\rm apo}$ are the median pericenter and apocenter of the fitted progenitor orbit, integrated backward in the time-dependent MW+LMC potential; $L$ is the angular extent of the \textit{Gaia} members, and the corresponding physical length at $d_\odot$. $N_{\rm Gaia}$ is the number of \textit{Gaia} members in the \cite{Bonaca2025} catalog; $N_{G<21}$ is the number of stars brighter than $G=21$ implied by $M_{\rm prog}$ for a 13~Gyr population at $d_\mathrm{\odot}$. $N_{\rm enc}$ is the number of subhalo encounters per stream above the quoted peak along-stream velocity kick $v_{\rm kick}$: mean over the 256 realizations, with the 16--84\% range.}
\end{deluxetable*}

Our simulation paints on a population of LMC subhalos ``on-rails'', modeling the flux of subhalos brought in by the LMC while ignoring the LMC-induced deformation and reflex motion of the MW's own subhalo population. 
We find that direct impacts with LMC subhalos are quite rare in our simulation: only 300S and Jet have a mean number of $\vkick > 0.1$\,\kms{} encounters with LMC subhalos $> 0.5$ across realizations. 
Due to its location on-sky, 300S receives a relatively high relative proportion ($\approx 30\%$) of its encounters from LMC subhalos, matching expectations from \cite{Arora2024}. 
A more detailed investigation of the LMC's influence will require a forward model of the MW and LMC's subhalo populations that incorporates the shared gravitational field of the systems. 
This would self-consistently capture the reflex response of the MW and its subhalos to the infalling LMC, and the associated anisotropy in the encounter rates of subhalos \citep[e.g.,][]{Erkal2021, Petersen2021, Arora2024, Chandra2025}.

\subsection{Properties of Stream-Encountering Subhalos}\label{sec:results_subhalos}

Quantifying stream--subhalo interactions with a cosmologically realistic subhalo model allows us to study the properties of the subhalo population that the MW's stellar streams are sensitive to. 
For this section, we restrict ourselves to the top 10 streams in terms of the total expected number of subhalo encounters (Figure~\ref{fig:ndetect}), and continue to adopt $v_\mathrm{kick} > 0.1$\,\kms{} as the regime of plausibly detectable encounters. 

\begin{figure}
    \centering 
    \includegraphics[width=\linewidth]{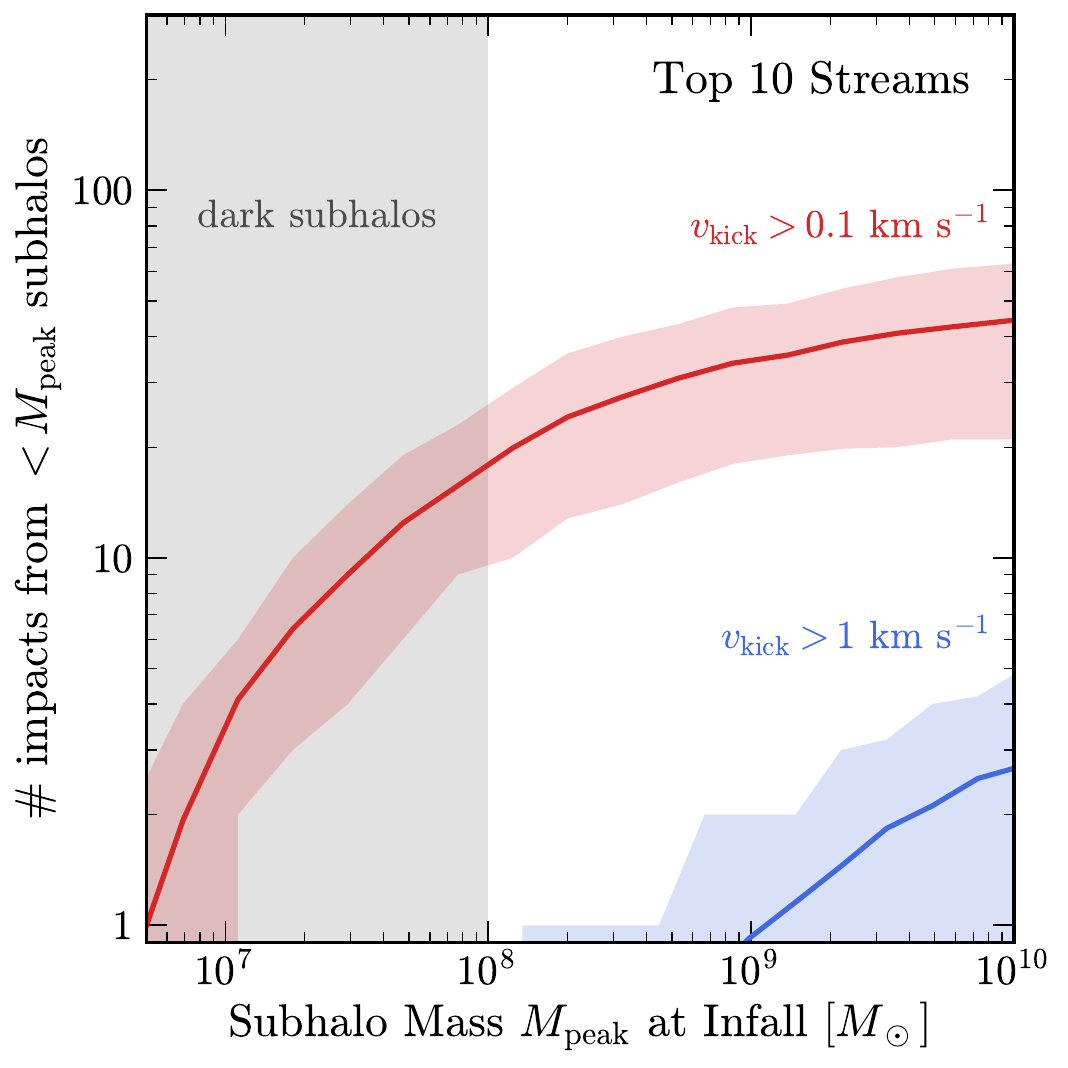}
    \caption{\textbf{Cumulative mass spectrum at infall of subhalos detectable by MW streams.} Restricting to the 10 streams with the largest expected encounter rate, the number of impacts from subhalos less massive than a given mass is shown, for two choices of the $v_\mathrm{kick}$ detection threshold. 
    The shaded region with $M_\mathrm{peak} < 10^8\,M_\odot$ corresponds to low-mass subhalos that are unlikely to host detectable dwarf galaxies. 
    }
    \label{fig:nimpact_cdf}
\end{figure}

One of the key considerations for streams as dark matter detectors is the prospect that they can reveal the influence of truly starless subhalos, below the threshold of dwarf galaxy formation. 
Not only does this reveal a mass scale of subhalos inaccessible to dwarf galaxy measurements, but it also enables us to study subhalo properties unaffected by baryonic processes in those galaxies. 
Although the physics of galaxy formation at the lowest-mass end is quite uncertain, a subhalo's likelihood of forming stars is primarily mediated by its peak mass $M_\mathrm{peak}$ prior to infall and subsequent tidal stripping \citep[e.g.,][]{Wechsler2018, Nadler2020, Nadler2024, Nadler2025}. 
The \gal{} subhalos encountered by MW streams have median bound mass fraction $\mu \equiv M_\mathrm{sub}/M_\mathrm{peak} \approx 6\%$, with a $16$--$84$th percentile range of $1$--$20\%$. 

Figure~\ref{fig:nimpact_cdf} shows the cumulative peak mass $M_\mathrm{peak}$ spectrum of subhalos that induce $\vkick > 1$\,\kms{} and $\vkick > 0.1$\,\kms{} perturbations to the top 10 MW streams. 
$\vkick > 1$\,\kms{} impacts are predominantly produced by subhalos with $M_\mathrm{peak} > 10^8\,\msun$, a mass scale at which ultra-faint dwarf galaxies are already thought to reside \citep{Nadler2024}. 
Although stream--subhalo detections would still be valuable in this regime, they would not probe a qualitatively different subhalo mass scale compared to observations of ultra-faint dwarf galaxies. 
On the other hand, a sensitivity to $v_\mathrm{kick} > 0.1$\,\kms{} impacts can reveal an order of magnitude more stream--subhalo encounters, including $\approx 15$ from $M_\mathrm{peak} \lesssim 10^8\,\msun{}$ subhalos that are unlikely to host galaxies. 
Constraining the MW's subhalo mass function in this regime can help distinguish competing dark matter models that dwarfs are insensitive to \citep[e.g.,][]{Nadler2021, Nibauer2026}.

\begin{figure}
    \centering
    \includegraphics[width=\linewidth]{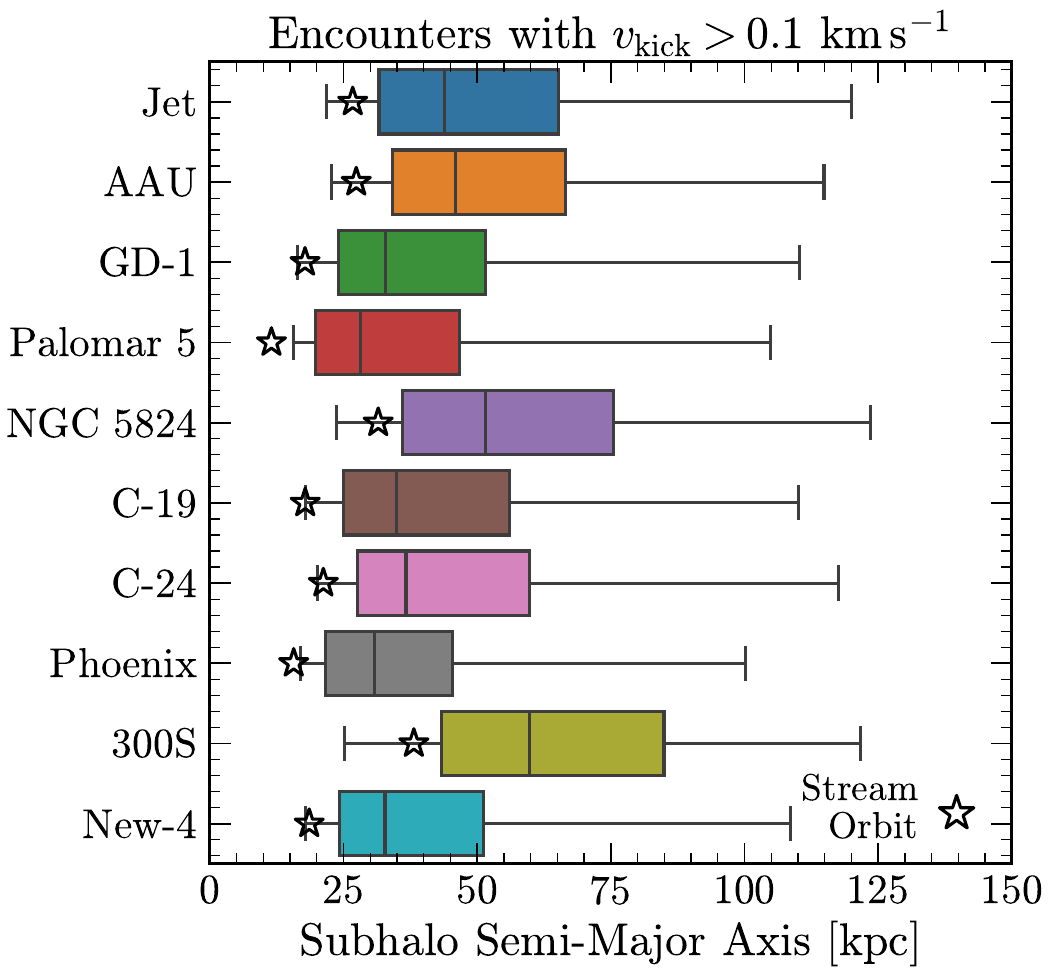}
    \caption{\textbf{Radial distribution of subhalos that encounter MW streams.} The semi-major axis of each subhalo is computed over its past orbit, and the corresponding value for each stream progenitor is indicated with a star. 
    Boxes indicate the median and inter-quartile range of all encounters across all realizations, and whiskers contain $90\%$ of encounters. 
    Observations of MW streams are most sensitive to subhalos residing $30$--$60$~kpc from the Galactic center.}
    \label{fig:rgal_dist}
\end{figure}

Connecting stream observations to the underlying DM model requires an understanding of where the stream-detectable subhalos reside, since that sets the range over which measured subhalo abundances can be compared to simulations. 
Furthermore, different DM models may alter the radial distributions of subhalos, although this effect may be small compared to the typical host-to-host scatter \citep[e.g.,][]{Nadler2025a}.
Figure~\ref{fig:rgal_dist} summarizes the radial orbits of subhalos that induce $\vkick > 0.1$\,\kms{} encounters with the top 10 MW streams.
The subhalo semi-major axis is computed over the most recent radial period, and the corresponding value for stream progenitors is shown as well.
Although our considered streams span a range of Galactocentric radii, the majority of subhalos encountered by MW streams reside well beyond the orbital extent of the streams themselves. 
To first order, this is a consequence of the much larger spatial volume at larger Galactocentric radii, coupled with the relatively eccentric distribution of \gal{} subhalos (median $e\sim 0.5$), which results in this being the dominant reservoir from which impacting subhalos are sourced.
A contributing effect is tidal stripping from the MW's disk, which suppresses the number density of subhalos at $r_\mathrm{gal} \lesssim 20$~kpc by up to $50\%$ (Figure~\ref{fig:nsub_disk}). 
Therefore, even relatively nearby streams are sensitive to subhalos residing far in the outer halo of the Galaxy. 

\begin{figure}
    \centering
    \includegraphics[width=\linewidth]{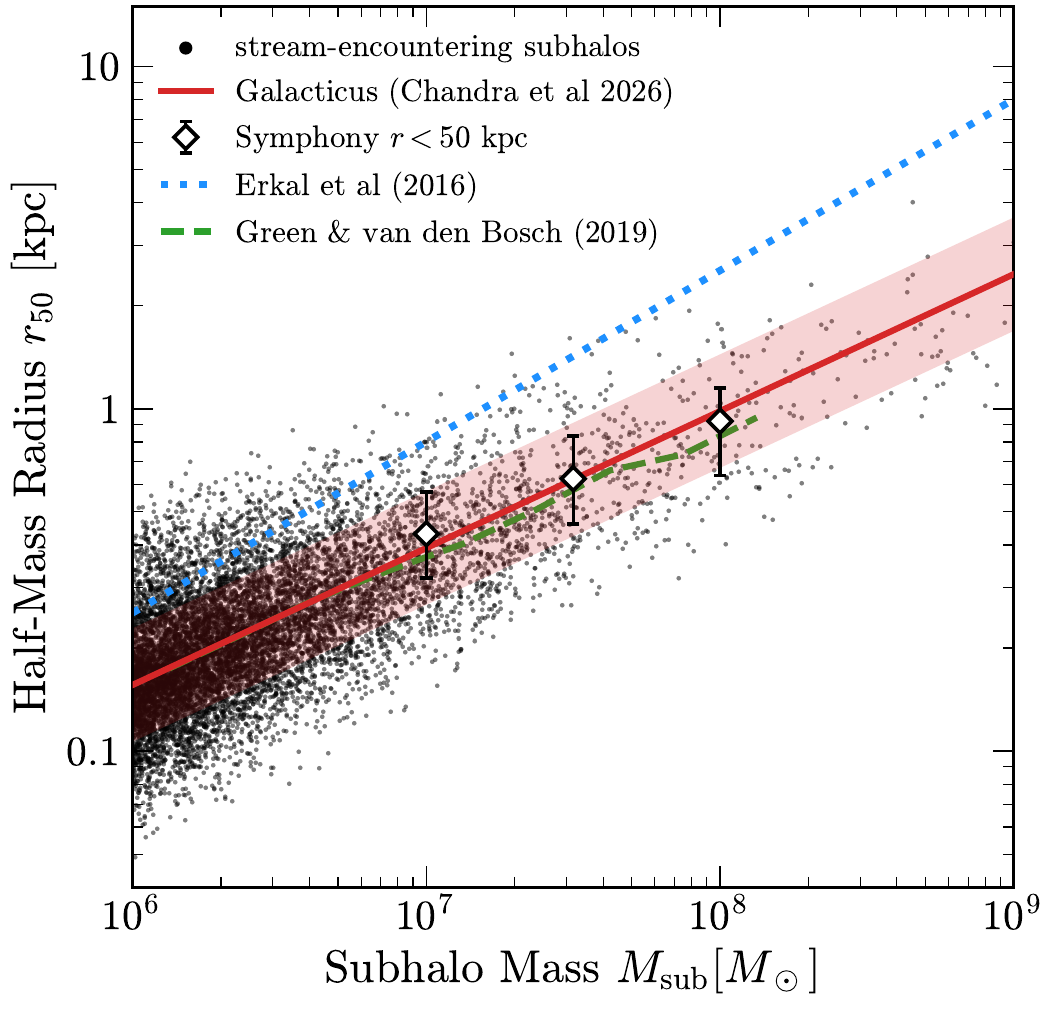}
    \caption{\textbf{Mass--size relation for \gal{} subhalos that encounter MW streams}. Points show unique subhalos across all our simulations that pass within 1~kpc of a MW stream; the red line shows our fitted relation. For comparison, we show the Hernquist mass--size relation assumed by \cite{Erkal2016a}, predictions from the tidal tracks of \cite{Green2019}, and measurements from the high-resolution runs of the \texttt{Symphony} $N$-body simulation \citep{Nadler2023, Mansfield2024}.}
    \label{fig:sizemass}
\end{figure}

Observations of streams are sensitive to the mass--size relation of dark subhalos, and measuring this relation can differentiate particle models of dark matter \citep[e.g.,][]{Nibauer2026}. 
Figure~\ref{fig:sizemass} shows the mass--size relation for all \gal{} subhalos that pass within $1$~kpc of any MW stream over the course of our simulation. 
For comparison, we show the median relation predicted by the tidal tracks of \cite{Green2019}, utilizing the infall masses, scale radii, and bound mass fractions of our \gal{} subhalos. 
We also show the relation assumed by the stream--subhalo forecasts of \cite{Erkal2016a} (and by extension, \citealt{Lu2026b}), which were fit to all subhalos around a MW-like simulated galaxy from the Via Lactea II simulations \citep[VLII;][]{Diemand2008}. 
Finally, we show measurements from the \texttt{Symphony} $N$-body simulation's \texttt{Symfind} subhalo catalogs \citep{Nadler2023, Mansfield2024}, using inner halo $r < 50$~kpc subhalos with $M_\mathrm{sub} > 5 \times 10^6\,M_\odot$ from the high-resolution `HR' runs of the simulation.  

Figure~\ref{fig:sizemass} illustrates that stream-encountering subhalos are $\approx 2\times$ more compact than the VLII relations fit by \cite{Erkal2016a}. 
Stream-encountering subhalos have orbits bringing them within $\lesssim 30$~kpc of the MW, resulting in a severe amount of tidal stripping that makes their profiles more compact. 
\cite{Erkal2016a} and \cite{Lu2026b} fit all VLII subhalos within the host's virial radius, weighting the mass--size relation towards the far more numerous outer-halo subhalos that are minimally stripped, but would almost never encounter MW streams.
The \gal{} mass--size relation of stream-encountering subhalos is consistent with the $r < 50$~kpc relation from the \texttt{Symphony} $N$-body simulations. 

We recommend that future works predicting the influence of CDM subhalos on streams use self-consistent stripped profiles calibrated to semi-analytic and $N$-body simulations \citep[e.g.,][]{Green2019, Gilman2020, Dacunha2026}. 
If an analytic profile is desired for computational simplicity, the mass--size relation predicted by our \gal{} simulation is \begin{equation}\label{eq:sizemass}
    r_{50} = 0.4~\text{kpc} \left( \frac{M_\mathrm{sub}}{10^7~\mathrm{M_\odot}} \right)^{0.4}
\end{equation} 
with an intrinsic log-normal scatter of
$\sigma_{\ln r_{50}} = 0.4$ ($0.17$~dex).
A population of perturbers with \cite{Hernquist1990} profiles can be constructed for any arbitrary subhalo mass function by setting the Hernquist subhalo scale radii $r_\mathrm{s,H}$ to \begin{equation}\label{eq:hernquist_a}
\begin{aligned}
    r_\mathrm{s,H}(M_\mathrm{sub}) &= \frac{f_\mathrm{scatter}\, r_{50}(M_\mathrm{sub})}{1+\sqrt{2}} \\
    f_\mathrm{scatter} &= e^{\epsilon}, \qquad \epsilon \sim \mathcal{N}\!\left(0,\, 0.4^{2}\right),
\end{aligned}
\end{equation}
where $r_{50}(M_\mathrm{sub})$ is given by Equation~\ref{eq:sizemass}.

We emphasize that the relation presented in Equation~\ref{eq:sizemass} only applies to the heavily stripped inner-halo subhalos that are likely to encounter MW streams.
Furthermore, these predictions will depend on the adopted prescription for subhalo tidal stripping. 
The \gal{} prescriptions used in this work are tuned to the Caterpillar cosmological $N$-body simulations \citep{Griffen2016}, but alternative prescriptions tuned to idealized simulations may predict a different degree of tidal stripping \citep[e.g.,][]{Du2024, Du2025}. 
Different tidal stripping recipes will alter the abundance of inner subhalos, and also affect the infall time and size distribution of surviving subhalos, influencing their internal structure at the present day. 
It will be valuable for future work to re-examine the \gal{} tidal stripping calibration with both idealized and cosmological simulations, in order to robustly predict the internal structure of stream-encountering subhalos \citep{Dacunha2026}. 

Adopting the physically motivated subhalo density profiles described here can have significant consequences for stream-based measurements of the subhalo mass function and mass--size relation. 
For example, \cite{Nibauer2026} use the velocity dispersion of GD-1 and the \cite{Erkal2016a} Hernquist subhalo model to infer that subhalos $10\times$ more compact than CDM are preferred at $M_\mathrm{sub} \approx 10^7\,M_\odot$. 
Figure~\ref{fig:sizemass} shows that the \gal{} stripped subhalo population is $\approx 2\times$ more compact than the \cite{Erkal2016a} model at that subhalo mass, loosening the GD-1 velocity dispersion's preference to $\approx 5\times$ more compact subhalos than CDM at $M_\mathrm{sub} \approx 10^7\,M_\odot$. 
Tidal stripping could therefore strongly influence the apparent tension of recent stream observations with CDM, motivating more detailed observations and modeling of streams like GD-1 with realistic subhalo populations. 

\subsection{Distinguishing Dark Matter Models}

What can the MW's stellar streams teach us about the underlying physics of dark matter? 
Our work quantifies stream--subhalo encounters under the assumption of the cold dark matter (CDM) model. 
These encounter statistics alone provide a useful avenue to explore what the MW streams can teach us about DM. 
A strength of the \gal{} semi-analytic model is that it can self-consistently incorporate alternative dark matter models, producing even higher-fidelity forecasts in alternative DM models.
We leave this detailed treatment to future work (see $\S$\ref{sec:future}), and adopt a simplified approach to estimate the effect of alternative DM models in this section.

Figure~\ref{fig:nimpact_cdf} demonstrates that sub-\kms{} perturbations reveal the influence of subhalos with peak masses plausibly below the threshold of galaxy formation. 
This is the regime that can test WDM, and many other models that suppress small-scale power. WDM, as typically implemented in small-scale structure probes, describes any model that has a free-streaming length scale which suppresses the formation of small-scale subhalos, encoded by $M_\mathrm{hm}$, the half-mode mass at which the linear matter transfer function is suppressed by $50\%$ due to WDM free-streaming. 
Depending on the exact physical model, $M_\mathrm{hm}$ maps to $m_\mathrm{WDM}$, the mass of the WDM particle. 
The current census of satellite galaxies around the MW excludes $M_\mathrm{hm} \gtrsim 4 \times 10^7\,M_\odot$ (corresponding to particle mass $m_\mathrm{WDM} \lesssim 6$~keV) at $95\%$ confidence \citep{Nadler2021, Nadler2025a}. 

To investigate the ability of MW streams to improve constraints on WDM models, we perform a simple post-processed resampling of our encounter catalog to emulate the effect of a SHMF suppression induced by WDM. 
$N$-body simulations suggest that such an approximation is adequate for our present purposes, since WDM predominantly influences the SHMF.
WDM subhalos are predicted to be slightly less compact than CDM subhalos at fixed mass, which would make the difference in stream perturbations between CDM and WDM more pronounced \citep{Bose2016, Ludlow2016}. 
We adopt the WDM processing of the present-day CDM mass function from \cite{Lovell2020}:
\begin{equation}
    \frac{n_\mathrm{WDM}}{n_\mathrm{CDM}}\left(M\right)
    = \left[1 + \left(\alpha\,\frac{M_\mathrm{hm}}{M}\right)^{\beta}\right]^{-\gamma}
    \label{eq:wdm_suppression}
\end{equation}
where $n_\mathrm{WDM}/n_\mathrm{CDM}$ is the ratio of the WDM to CDM subhalo mass functions, $M_\mathrm{hm}$ is the half-mode mass, and $(\alpha, \beta, \gamma) = (4.2,\, 2.5,\, 0.2)$ are the subhalo-calibrated
coefficients of \citet{Lovell2020}. 
Equation~\ref{eq:wdm_suppression} defines a weighting function that we can use to adjust any of our CDM-based counts for a given WDM model, parameterized by the half-mode mass $M_\mathrm{hm}$. 

\begin{figure}
    \centering
    \includegraphics[width=\linewidth]{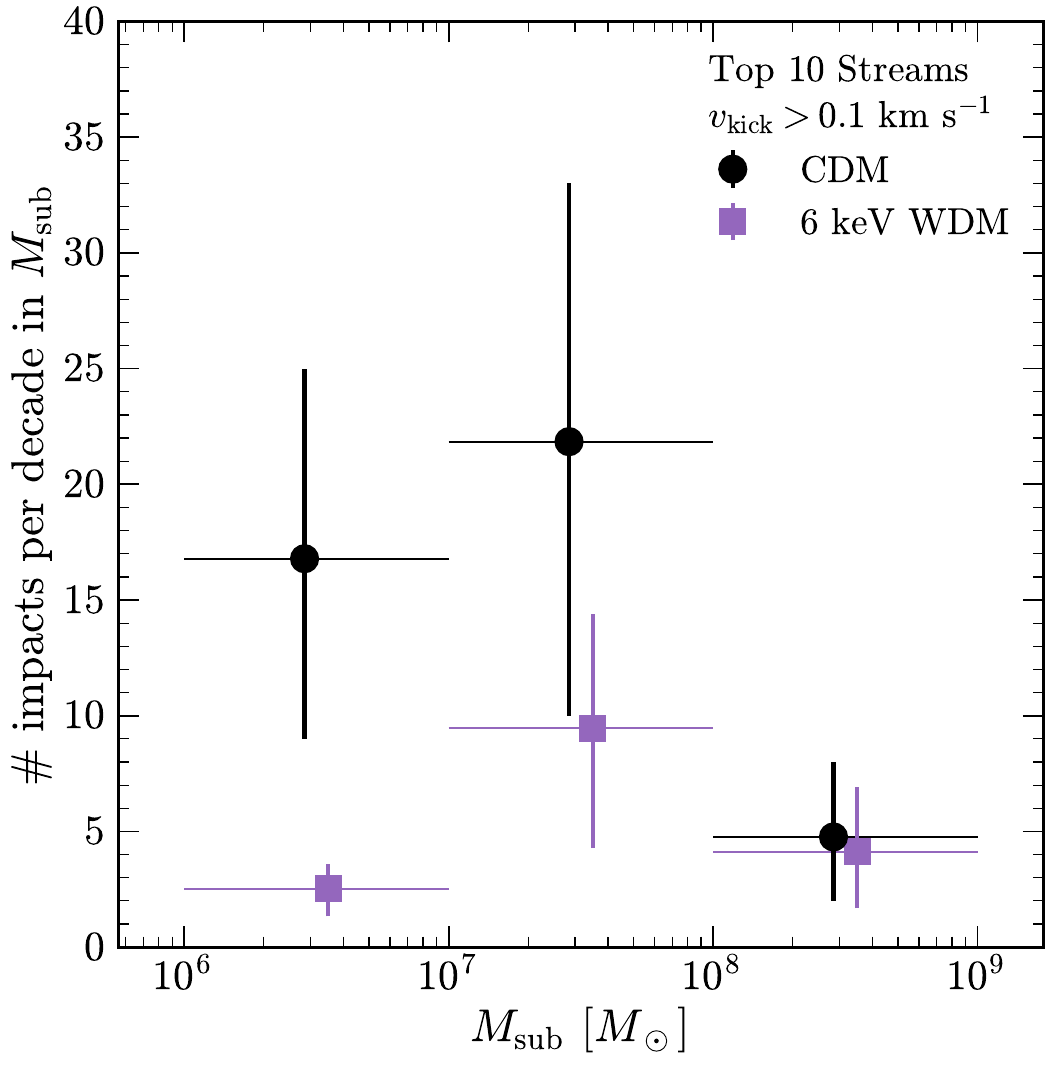}
    \caption{\textbf{The mass spectrum of encounters predicted from cold and warm dark matter models.}
    The WDM model has a half-mode mass $M_\mathrm{hm} = 4 \times 10^7\,M_\odot$, corresponding to $m_\mathrm{WDM} \approx 6$~keV.
    %The cumulative distribution of $\vkick$ encounters with the top 10 MW streams in CDM and WDM respectively, where the WDM prediction is calculated by down-sampling the CDM predictions from our simulation. \textbf{Bottom:} 
    The number of $\vkick > 0.1$\,\kms{} encounters shown is aggregated across the top 10 MW streams per decade in the subhalo mass $M_\mathrm{sub}$ at the encounter time, with bars showing the $1\sigma$ scatter across realizations.}
    \label{fig:wdm}
\end{figure}

Figure~\ref{fig:wdm} shows predictions from resampling our simulation to the WDM prediction for $M_\mathrm{hm} = 4 \times 10^7\,M_\odot$ ($m_\mathrm{WDM} \approx 6$~keV; \citealt{Viel2005, Schneider2012}), the warmest model (lightest WDM particle) allowed by measurements of satellite galaxies \citep{Nadler2021, Nadler2025a}. 
The mass spectrum of subhalo masses at the time of impact is shown for $\vkick > 0.1$\,\kms{} encounters across the top 10 streams.
This figure highlights the strong suppression of potentially resolvable encounters with $M_\mathrm{sub} \lesssim 10^7\,M_\odot$ subhalos. 
Identifying $\gtrsim 5\text{--}10$ individual subhalo encounters from subhalos at those mass scales would put strong constraints on the WDM particle mass. 
A key uncertainty in differentiating CDM and WDM models is the intrinsic scatter in the underlying subhalo population, since our observations in the MW naturally only give us a single realization of the measurement. 
Our \gal{} simulation has a between-tree scatter (in terms of the number of stream--subhalo encounters) of $\approx 40\%$, driven by variations in the number density of subhalos in the inner regions of the Galaxy probed by streams. 
Complementary external priors on this underlying richness of subhalos---for example from a census of satellite galaxies---will help strengthen the interpretation of stream perturbations \citep[e.g.,][]{Nadler2021, Banik2021streams}.

\begin{figure}
    \centering
    \includegraphics[width=\linewidth]{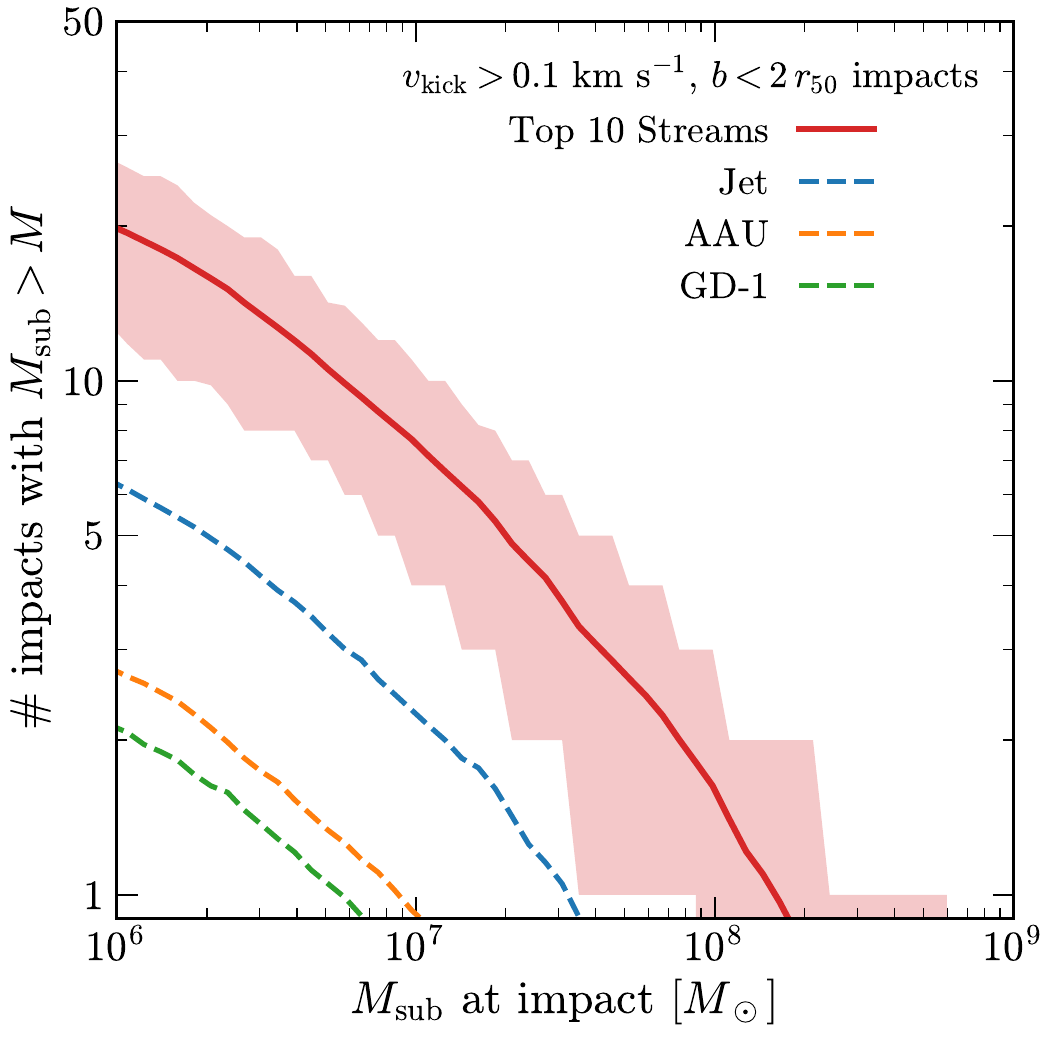}
    \caption{\textbf{Number of close stream--subhalo encounters that could measure the internal density structure of subhalos.} 
    The cumulative distribution of significant encounters with impact parameters smaller than twice the subhalo half-mass radius is shown, aggregated across the top 10 MW streams (with 1$\sigma$ scatter between realizations shown) and individually for the top 3 streams. }
    \label{fig:nclose}
\end{figure}

Stream--subhalo encounters can also test models of self-interacting dark matter (SIDM; \citealt{Spergel2000}), which produce core-collapsed subhalos with substantially higher densities than CDM predicts. 
This results in stronger perturbation signatures for close encounters where the impact parameter is comparable to the size of the subhalo $b \lesssim 2\cdot r_\mathrm{50}$, where $r_\mathrm{50}$ is the half-mass radius of the subhalo. 
More distant encounters reduce to the point-mass limit, and are therefore insensitive to the density profile of subhalos.
Figure~\ref{fig:nclose} shows the number of close encounters with $\vkick > 0.1$\,\kms{} and impact parameters $b \lesssim 2\cdot r_\mathrm{50}$, aggregated across the top 10 MW streams, and individually for the top 3 MW streams. 
The perturbations from these close encounters would differ substantially depending on the internal density profiles of the subhalos, for example becoming larger and more localized for a dense, core-collapsed subhalo. 
The individual and integrated velocity perturbations of these streams could therefore be sensitive to differences between CDM and SIDM \citep[e.g.,][]{Zhang2025, Nibauer2026}.
Jet is a particularly promising case with $\gtrsim 5$ expected close encounters with $> 10^6\,M_\odot$ subhalos. 

\section{Discussion}\label{sec:discuss}

\subsection{Detectable Signatures of Dark Subhalos}\label{sec:detectable}

\begin{figure}
    \centering
    \includegraphics[width=\linewidth]{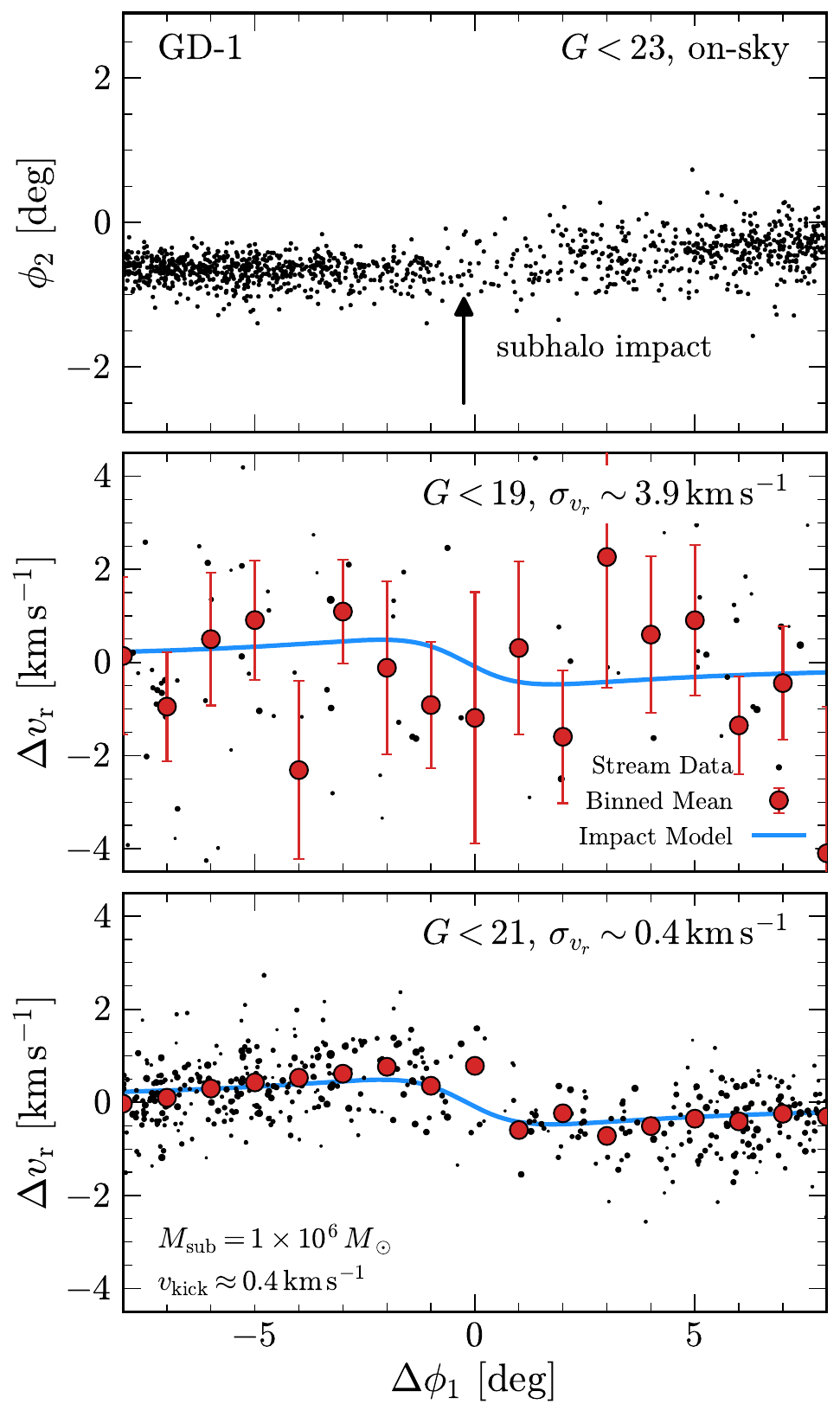}
    \caption{\textbf{Simulation of the GD-1 stream after encountering a $10^{6}\,\msun$ subhalo.} The top panel shows the on-sky density structure of $G < 23$ stars. The middle panel shows RV observations analogous to current low-resolution surveys, and the bottom panel shows high-resolution observations analogous to those planned by the Via Survey.
    Black points show individual observations---with larger marker sizes corresponding to brighter stars, and the median RV error of the plotted stars indicated in the label---and red points indicate the binned mean trend. 
    }
    \label{fig:impact_sim}
\end{figure}

So far, we have used the $\vkick$ statistic to summarize the interaction strength between MW streams and \gal{} subhalos. 
In this section, we briefly consider the observable signatures of these impacts.
Figure~\ref{fig:impact_sim} provides an illustrative example of a single subhalo encounter with the GD-1 stream, as predicted by our simulation. 
We use the \texttt{streamsculptor} code \citep{Nibauer2025} to generate the perturbed stream model, using the stream initial conditions and subhalo orbital properties for a single encounter from our simulation: a $\vkick \approx 0.4$\,\kms{} encounter $\approx 800$~Myr ago with a $M_\mathrm{sub} \approx 10^6\,M_\odot$ subhalo.
The intrinsic velocity dispersion predicted by the \cite{Chen2025} particle spray model is $\approx 1$~\kms{}, smaller than the measured dispersion of GD-1 from current surveys, since the measured dispersion may itself include perturbations from subhalos \citep{Jarvis2026, Nibauer2026}. 
The top panel shows the on-sky density of stars down to $G < 23$.
The middle panel shows mock RV observations analogous to low-resolution spectroscopic surveys like DESI, adopting magnitude-dependent RV error curves from \cite{Koposov2026} down to $G < 19$. 
The bottom panel shows mock RV observations from the instrument model for the Via Survey \citep{ViaCollaboration2026} down to $G < 21$, illustrating the ability to individually resolve sub-\kms{} $\vkick$ encounters with upcoming data.
Spectroscopic membership information can also help to detect and confirm the gap formed by this low-mass subhalo encounter, a regime in which photometry alone may be insufficient \citep[e.g.,][]{Pelissier2026, Boone2026}. 

\cite{Hilmi2024} and \cite{NGUYEN2025} are two other recent works that forecast the detectability of individual subhalo encounters, focusing on a handful of stream--subhalo examples. 
\cite{NGUYEN2025} find that incorporating RV information has a minimal influence on the ability to measure the mass of a $10^7\,M_\odot$ perturber, compared to density and proper motion data alone.  
Conversely, \cite{AnauMontel2026} measure the `diffusive' influence of many weak encounters and find that incorporating RV information improves the constraints on WDM models substantially. 

\cite{Lu2026} and \cite{Lu2026b} perform an investigation of the detectability of individual subhalo encounters with MW streams, forward-modeling the perturbations' influence on stream stars with analytic models and fitting idealized mock observations with various assumed instrument properties. 
\cite{Lu2026} find that on-sky information from \textit{Gaia} alone limits the minimum detectable subhalo mass to $M_\mathrm{sub} \gtrsim 10^7\,M_\odot$ for most streams. 
However, \cite{Lu2026} find that incorporating LSST proper motions and radial velocities from an idealized instrument model for the Via Survey \citep{Ivezic2019, ViaCollaboration2026} reduces the minimum detectable subhalo mass to $\gtrsim 10^6\,M_\odot$ for many of the top streams. 
In particular, several of the streams we predict to be maximally perturbed by subhalos (Palomar 5, AAU, GD-1) lie near the top of the list in terms of their fidelity as DM detectors. 

\cite{Lu2026b} extend the work of \cite{Lu2026} by adopting the \cite{Erkal2016a} analytic model (with no disk depletion) for the expected number of subhalo encounters each stream experiences, and find that Jet, AAU, Palomar~5, and GD-1 are streams with the highest likelihood of having individually detectable encounters with subhalos. 
Orphan-Chenab also ranks highly in their list, but we do not consider it in our work due to its large physical width and likely origin as a dwarf galaxy rather than a globular cluster \citep{Koposov2019, Li2022b}.  
Our disk-depleted \gal{} simulation predicts a comparable number of stream-encountering subhalos as the un-depleted \cite{Erkal2016a} model used by \cite{Lu2026b}. 
We also predict that individual close encounters will be stronger, given the more compact mass--size relation of \gal{} subhalos compared to the \cite{Erkal2016a} Hernquist model used by \cite{Lu2026b}. 
Our fitted dynamical ages agree with those of \cite{Lu2026b} for Jet and AAU, but are $\approx 1.5$--$2\times$ older for GD-1, Palomar~5, and 300S, likely because we adopt the $N$-body-calibrated \cite{Chen2025} particle-spray algorithm rather than \cite{Fardal2015}.

The detectability model adopted by \cite{Lu2026b} is likely pessimistic in terms of the information content provided by RV observations, preferentially selecting the strongest encounters that create gaps and distort the stream track.
The observability statistic adopted by \cite{Lu2026b} is primarily driven by morphological (density and width) information (see Figure~9 in \citealt{Lu2026b}). 
The reason for this morphological preference is that the baseline observational model in \citet{Lu2026b} assumes perfect stream membership information down to the 10-year LSST \citep{Ivezic2019} depth of $r < 27$, and assumes radial velocity uncertainties from the Via Survey's instrument model \citep{ViaCollaboration2026} down to $G < 24$. 
In practice, the depth of the Via Survey is $G < 21$, with typical statistical RV errors rising to $\gg 1$\,\kms{} fainter than $G > 21$ in the one-hour survey configuration.
Therefore, the vast majority of stars in the \cite{Lu2026b} mock data from $G \gtrsim 22$ to $r < 27$ have perfect positional information and essentially no RV information. 
Stream morphological information will be substantially degraded fainter than $r > 23$ due to contamination from the MW background (in the absence of proper motion information) and eventually due to contamination from galaxies \citep{Pelissier2026}. 

The integrated effect of subhalo impacts also distorts the velocity track of stream stars, leading to an increased RV dispersion \citep[e.g.,][]{Carlberg2024, Carlberg2025, Nibauer2025}. 
\cite{Nibauer2026} analyze the velocity dispersion of GD-1 and derive constraints on the abundance of low-mass subhalos and their mass--size relation, finding that the RV dispersion is a sensitive probe of the concentration of subhalos. 
It will be valuable to refine these various detectability frameworks further with realistic sample selections, in order to map the encounter rates predicted by our work to mock observations. 
Such calculations can provide a more accurate sense of which subhalo encounters may be individually detectable by future surveys, and which encounters can be inferred in aggregate through ensemble statistics like velocity dispersions and power spectra. 

\subsection{Future Work}\label{sec:future}

Our work has forecasted the population of \gal{} subhalos assuming a CDM model, the gravitational influence of the MW disk, and a simplified flux of incident subhalos from the LMC. 
Several avenues of work remain in order to robustly connect observations of stellar streams to physical models of DM: 

\paragraph{MW-matched subhalos with varying DM physics} As Figure~\ref{fig:nsub_comparison} illustrates, systematic discrepancies persist in predicting the abundance of MW subhalos even in the CDM model. 
Diagnosing and calibrating these uncertainties will require a mixture of full numerical simulations and semi-analytic models, for example to quantify the expected tidal stripping of low-mass subhalos in the inner $\lesssim 100$~kpc of MW-like host halos \citep[e.g.,][]{Dacunha2026}. 
A key strength of the \gal{} framework is that different assumptions can be directly varied in the underlying physical model, in order to produce self-consistent predictions for alternative physical models. 
For example, different DM physics can be implemented in \gal{} to model alterations to the SHMF and subhalo density profiles \citep[e.g.,][]{Nadler2025ddm, Lonergan2026, Delos2026b, Qin2026}. 
Furthermore, the underlying merger trees can be down-selected to pick closer analogs to the MW's own merger history, including the infalling LMC \citep[e.g.,][]{Nadler2023trees, Buch2024}. 
More work would be required to accurately model the influence of the LMC on the subhalo population of the MW+LMC system, in order to test and predict LMC-driven spatial anisotropies in the encounter rates of streams \citep[e.g.,][]{Arora2024}. 
Once semi-analytic models like \gal{} can produce robust statistical realizations of subhalo populations in different DM models, these population predictions can be folded into rapid inference frameworks to improve the fidelity of DM constraints \citep[e.g.,][]{Nibauer2025}. 

\paragraph{Stream initial conditions} Although we have constructed realistic initial conditions for the MW streams considered in this work, the quality of our stream simulations is ultimately limited by the underlying data of member stars. 
Whereas some streams have exceptionally clean \textit{Gaia} data and follow-up radial velocities (Figure~\ref{fig:streamfit}), many streams have patchy/incomplete membership data \citep{Bonaca2025}. 
Indeed, several \textit{Gaia}-discovered streams lack any RV information, which is required to confirm their nature as streams, and to model their progenitor initial conditions. 
Wide-field surveys like the DESI Milky Way Survey \citep{DESICollaboration2016, Cooper2023, Koposov2026}, the SDSS-V Milky Way Mapper Halo Survey \citep{Kollmeier2025, Chandra2026}, and the Via Survey \citep{ViaCollaboration2026} will be invaluable to refine the identity and membership of candidate MW streams. 
Furthermore, additional datasets like RR Lyrae stars can be used to better measure distances to streams \citep{DOMINGUEZ2026}.
Finally, several streams have surviving progenitors, whose stellar masses and orbits can be used to refine the stream models further. 

\paragraph{Influence of stream progenitors} The properties of the stream's progenitor clusters and their tidal release mechanisms strongly influence the structure of stellar streams. 
For example, the \cite{Chen2025} particle spray algorithm adopted in this work predicts stream dynamical ages $\approx 50\%$ older than the commonly used \cite{Fardal2015} algorithm. 
Accurately predicting the intrinsic widths and velocity dispersions of streams will require particle-spray algorithms that are tuned to predictions from $N$-body simulations, including important effects like variable mass loss, dynamical ejection of stars, and the effect of binaries, mass segregation, and stellar evolution on which stream stars are visible to surveys \citep[e.g.,][]{Kupper2008, Kupper2011, Kupper2012, Weatherford2023, Weatherford2024, Weatherford2026}. 
The abundance and distribution of binary stars in streams can also confound RV observations, requiring multi-epoch observations to detect and model the effect of binaries \citep[e.g.,][]{Kouwenhoven2008, Gieles2010, Badenes2018, Pianta2022, Wirth2024, OConnor2026, Phillips2026}.

\paragraph{Influence of the time-varying MW potential}  
Beyond forecasting encounter rates, realistic stream simulations are essential to predict the intrinsic structure of each stream in the absence of perturbations from DM subhalos. 
A key ingredient is the underlying gravitational potential of the MW, which is expected to have several non-axisymmetric and/or time-varying components like a triaxial DM halo, rotating bar, growing/tilting disk, giant molecular clouds, and the infalling LMC \citep[e.g.,][]{Erkal2016, Amorisco2016, HATTORI2016, Doke2022, Nibauer2024, Nibauer2025, Nibauer2025tilt, Panithanpaisal2026}. 
\cite{Arora2026} model streams in full hydrodynamical simulations, finding that their structure is substantially altered by the gravitational influence of the host galaxies. 
The strongest predictor of baryonic perturbations is unsurprisingly the orbital pericenter $r_\mathrm{peri}$. 
Several of the promising streams presented in Table~\ref{tab:streams} have pericenters $r_\mathrm{peri} > 10$~kpc, with Jet having a large $r_\mathrm{peri} \approx 17$~kpc, making it a particularly robust DM detector. 
It will be valuable for future work to develop simulations of stream evolution under controlled variations of the MW's individual potential components, and quantify which streams are most resistant to baryonic perturbations. 
It will also be valuable to investigate whether velocity information can disentangle baryonic influences from true encounters with subhalos. 
Finally, identifying additional distant streams like Jet will be valuable, particularly for future spectroscopic surveys that could efficiently observe streams beyond $> 30$~kpc \citep[e.g., Spec-S5,][]{Besuner2025}.

\paragraph{Spectroscopic detectability of subhalo encounters} This work has presented the rates of stream--subhalo encounters without assuming any particular observational detectability framework. 
It will be valuable for future work to perform a more detailed investigation on what factors affect the detectability of encounters above a certain $\vkick$ threshold, both individually and through aggregate statistics, both with morphology and with RV measurements \citep[e.g.,][]{Lu2026b, Pelissier2026}. 
These forecasts will complement upcoming all-sky spectroscopic surveys of streams like Via \citep{ViaCollaboration2026}, DESI Run 2 \citep{Schlegel2022}, and Spec-S5 \citep{Besuner2025}. 
Several factors should affect the detectability of an encounter beyond its overall strength. 
The orbital evolution of stream stars will cause older encounters to develop density and stream track features, and even the velocity signature of a given impact will evolve as stream stars oscillate around the stream track. 
Another important factor is the projection of a given stream segment along the current line of sight: a larger distance gradient along the line of sight will cause a larger component of the along-stream velocity kick to be directly visible in RV measurements. 
Directly modeling these various effects over the range of plausible encounter strengths and geometries will enable a forecast of the observational sensitivity of different stream regions to subhalo encounters, aiding survey planning and improving the fidelity of stream--subhalo inferences.

\section{Conclusions}\label{sec:conclusions}

In this work, we have constructed a semi-analytic simulation of the Milky Way's DM subhalo population, and fit particle-spray models for the evolution of $24$ long globular cluster streams. 
Quantifying the rates and properties of stream--subhalo encounters, we have identified several promising target streams for future spectroscopic surveys, and explored what these observations could teach us about the nature of DM around our Galaxy. 
Our main conclusions are as follows: 

\begin{itemize}
    \item We use the semi-analytic model \gal{} to predict the distribution of CDM subhalos around the Milky Way, including the effects of tidal stripping, the gravitational influence of the MW's disk, and subhalos contributed by the LMC. 
    Discrepancies at the factor-of-two level persist between semi-analytic prescriptions and $N$-body simulations, particularly in the inner regions of the galaxy 
    (Figure~\ref{fig:nsub_comparison}).
    Since stream--subhalo encounter rates scale linearly with the assumed number density of subhalos, the abundance of subhalos in the inner Galaxy remains a dominant uncertainty in connecting stream observations to physical models of DM.

    \item Selecting \nstream{} thin globular cluster streams from the catalog of \cite{Bonaca2025}, we derive their progenitor masses, dynamical ages, and orbital trajectories using a full particle-spray forward-model in a realistic MW potential that includes the LMC (Figures~\ref{fig:skymap}, \ref{fig:streamfit}, \ref{fig:streamfit_summary}). 
    
    \item We evaluate interactions between MW streams and \gal{} subhalos over the past 4~Gyr, cataloging close encounters and estimating their peak along-stream velocity perturbation $\vkick$ in the impulse approximation (Figure~\ref{fig:impactmovie}). In our direct simulation, the early stripped outer tails of the streams experience the largest number of encounters over the lifetime of the stream (Figure~\ref{fig:phi1_trel}). 
    The inner tails are more sensitive to the most recent encounters, and could be used to calibrate the intrinsic velocity dispersion of the stream. 
    
    \item Several streams stand out as having experienced substantial subhalo perturbations: Jet, AAU, GD-1, and Palomar~5 (Figure~\ref{fig:ndetect}). 
    NGC 5824, C-19, C-24, Phoenix, 300S, and New-4 round out the list of the top 10 streams in terms of predicted encounters. 
    The top MW streams undergo $2$--$12$ significant $\vkick > 0.1$\,\kms{} encounters each, with an individual $\approx 70\%$ scatter driven by Poisson fluctuations and host-host scatter in the abundance of subhalos that reach the inner galaxy.  
    
    \item Jet, AAU, GD-1, and C-24 are observationally favorable and exposed to fewer baryonic perturbations from the MW's disk ($\gtrsim 500$ observable stars down to the $G < 21$ \textit{Gaia} limit with orbital pericenters $r_\mathrm{peri} > 10$~kpc; Figure~\ref{fig:observability}).
    Jet is a remarkable outlier ($\approx 2\times$ the encounters of the next best stream) due to its old dynamical age and long physical extent, representing a promising target for future spectroscopic surveys. 
    Jet also has a large orbital pericenter ($r_\mathrm{peri} \approx 17$~kpc), making it relatively insensitive to baryonic perturbations from the MW's disk.   

    \item Our direct simulation predicts the structural and orbital properties of stream-detectable subhalos. 
    Given the radially anisotropic velocity distribution of subhalos, MW streams are sensitive to the entire reservoir of subhalos from $r_\mathrm{gal} \approx 30$--$60$~kpc (Figure~\ref{fig:rgal_dist}). 
    Stream-encountering subhalos have been tidally stripped to between $1$--$20\%$ of their peak mass at infall, and are consequently $\approx 2\times$ more compact than previous stream--subhalo models have assumed (Figure~\ref{fig:sizemass}). 
    The numerical profiles of individual subhalos predicted by \gal{} are well-fit by a truncated NFW model (Figure~\ref{fig:subdensity}). 
    We provide a model for the mass--size relation of stripped stream-encountering subhalos in Equation~\ref{eq:sizemass}. 
    Adopting a realistic density model for stripped subhalos---including the resulting subhalo mass--size relation and its intrinsic scatter---could partially alleviate the apparent tension between recent stream observations and CDM subhalos. 
    
    \item  The peak mass of each subhalo prior to tidal stripping predicts the likelihood of a subhalo to host a satellite galaxy. 
    A sensitivity to $\vkick > 0.1\,$\kms{} encounters can detect $\approx 15$ encounters from truly dark subhalos with peak masses $M_\mathrm{peak} < 10^8\,M_\odot$, below the threshold of galaxy formation (Figure~\ref{fig:nimpact_cdf}).
    Spectroscopic observations of stellar streams could therefore detect dark matter substructure below the scales probed by dwarf satellite galaxies. 
    
    \item Adopting a simple down-sampling scheme to emulate the suppression of low-mass subhalos predicted by WDM models, we find that the abundance of $\vkick \gtrsim 0.1$\,\kms{} encounters could differentiate these DM models at lower mass scales than dwarf galaxies currently do (Figure~\ref{fig:wdm}). 
    Furthermore, the streams presented here have a high likelihood of hosting several close $b < 2 \cdot r_\mathrm{50}$ subhalo encounters, which are sensitive to subhalo density profiles and could distinguish models of self-interacting dark matter (Figure~\ref{fig:nclose}).

    \item Although a full detectability forecast is left to future work, upcoming spectroscopic surveys could individually detect sub-\kms{} $\vkick$ perturbations, 
    and infer the abundance of smaller subhalo perturbations through integrated measurements like stream velocity dispersion profiles and power spectra (Figure~\ref{fig:impact_sim}). 
    Developing robust statistical estimators that leverage both density and velocity information will be vital to complement future surveys, and to realize the stream--subhalo encounter census forecasted in this work. 
    
\end{itemize}

\begin{acknowledgments}

We thank 
Ebtihal Abdelaziz,
Gianfranco Bertone, 
Andreas Burkert,
Ethan Nadler,
Matthieu Pelissier,
Nathaniel Starkman,
and Julie Xue
for conversations and feedback that improved this work. 
This work has benefited from conversations at the 2026 Probe Combination for Dark Matter Physics in the Era of Large Surveys conference at the University of Chicago, and from conversations started at the Kavli Institute for Theoretical Physics (KITP), supported by grant NSF PHY-2309135 to KITP.

Support for this work was provided by NASA through the NASA Hubble Fellowship grant HST-HF2-51600.001-A awarded by the Space Telescope Science Institute, which is operated by the Association of Universities for Research in Astronomy, Incorporated, under NASA contract NAS5-26555. RHW, TD, and PM received partial support from the Kavli Foundation for their work on the Via Project. 
The computations in this paper were run on the FASRC Cannon cluster supported by the FAS Division of Science Research Computing Group at Harvard University. This work used data from the Symphony suite of simulations (http://web.stanford.edu/group/gfc/symphony/), which was supported by the Kavli Institute for Particle Astrophysics and Cosmology at Stanford University and SLAC National Accelerator Laboratory, and by the U.S. Department of Energy under contract number DE-AC02-76SF00515 to SLAC National Accelerator Laboratory.
This research has made extensive use of NASA's Astrophysics Data System Bibliographic Services.
This work made use of \texttt{OverCite} \citep{Shariat2026}, an in-editor citation tool for \LaTeX.
Anthropic's Claude large language model (LLM) was used to review and revise the simulation and analysis code underlying this work, and to assist with proofreading, literature review, and auditing the internal consistency of the manuscript. 
No part of this manuscript was written by an LLM; suggested edits were reviewed and adopted by the authors at their discretion.
The authors take full responsibility for the content of this work.  

This work has made use of data from the European Space Agency (ESA) mission {\it Gaia} (\url{https://www.cosmos.esa.int/gaia}), processed by the {\it Gaia} Data Processing and Analysis Consortium (DPAC, \url{https://www.cosmos.esa.int/web/gaia/dpac/consortium}). Funding for the DPAC has been provided by national institutions, in particular the institutions participating in the {\it Gaia} Multilateral Agreement. 

\end{acknowledgments}

\software{\texttt{agama} \citep{Vasiliev2019},
\texttt{astropy} \citep{AstropyCollaboration2013, AstropyCollaboration2018, AstropyCollaboration2022},
\texttt{gala} \citep{gala},
\texttt{galacticus} \citep{Benson2012},
\texttt{matplotlib} \citep{Hunter2007}, 
\texttt{numpy} \citep{Harris2020}, 
\texttt{scipy} \citep{Virtanen2020}, 
\texttt{streamsculptor} \citep{Nibauer2025}
}

\facilities{Gaia}

\bibliography{chandra, bib}
\bibliographystyle{aasjournal}

\end{document}